# CHEMICAL CLASSIFICATION OF SPHERULES RECOVERED FROM THE PACIFIC OCEAN SITE OF THE CNEOS 2014-01-08 (IM1) BOLIDE.

A. Loeb[1,2], S.B. Jacobsen[2,3], R. Tagle[2], T. Adamson[2], S. Bergstrom[2], J. Cherston [1,2], R. Cloete[1,2], S. Cohen[2,7], L. Domine[1,2], H. Fu[2,3], C. Hoskinson[2], E. Hyung[2,3], M. Kelly[2], E. Lard[2], F. Laukien[2,6], J. Lem[2,5], R. McCallum[2], R. Millsap[2], C. Parendo[2,3], C. Peddeti[2,4], J. Pugh[2], S. Samuha[2,7], D.D. Sasselov[1,2], M. Schlereth[2], J. Siler[2], A. Siraj[1,2], P.M. Smith[2], J. Taylor[2], R. Weed[2,4], A. Wright[2], J. Wynn[2].

[1]Dept. of Astronomy, Harvard Univ., Cambridge, 02138, MA (aloeb@cfa.harvard.edu).
[2]Interstellar Expedition of the Galileo Project, Cambridge, 02138, MA.
[3]Dept. of Earth and Planet. Sci., Harvard Univ., Cambridge, 02138, MA.
[4]Dept. of Nuclear Eng., Univ. of California Berkeley, Berkeley, 94720, CA.
[5]Dept. of Mining Eng., PNG Univ. of Technology, Lae, 411, Papua New Guinea.
[6]Dept. of Chemistry and Chemical Biology, Harvard Univ., Cambridge, 02138, MA.
[7]Dept. of Materials Eng., NRCN, P.O. Box 9001, Beer-Sheva, 84190, Israel.

**Abstract:** We have conducted an extensive towed-magnetic-sled survey during the period of June 14-28, 2023, over the seafloor about 85 km north of Manus Island, Papua New Guinea, centered around the calculated path of the bolide CNEOS 2014-01-08 (IM1). We found about 850 spherules of diameter 0.1-1.3 millimeters in our samples. The samples were analyzed by micro-XRF, Electron Probe Microanalyzer and ICP Mass spectrometry. Here we report major and trace element compositions of the samples and classify spherules based on that analysis. We identified 78% of the spherules as primitive, in that their compositions have not been affected by planetary differentiation. We divided these into four groups, three of which correspond to previously described cosmic spherule types. Spherules in the fourth group, comprising 22% of the collection, appear to all reflect planetary igneous differentiation and are all different from previously described spherules. We call them D-type spherules. At least five of the D-type spherules are suggested to be terrestrial in origin, although many spherules exhibit elemental ratios that are distinct from known planetary bodies and their origins are undetermined. A subset of the D-spherules show an excess of Be, La and U, by up to three orders of magnitude relative to the solar system standard of CI chondrites. Detailed mass spectrometry of 12 of these "BeLaU"-type spherules, the population of which may constitute up to ~10% of our entire collected sample, suggests that they are derived from material formed by planetary igneous fractionation. Their chemical composition is unlike any known solar system material. We compare these compositions to known differentiated bodies in the solar system and find them similar to those of evolved planetary materials–with lunar KREEP the closest in terms of its trace element enrichment pattern, but unusual in terms of their elevated CI-normalized incompatible elements. The "BeLaU"-type spherules reflect a highly differentiated, extremely evolved composition of an unknown source.

## Introduction

The retrieval of cosmic spherules from meteor sites has a long history, with related morphology and composition analyses linking them to various components of the solar system (Brownlee et al., 1979; Maurette et al., 1991; Taylor and Brownlee, 1991; Xue et al., 1994; Brownlee et al., 1997; Herzog et al., 1999; Taylor et al., 2000; Engrand et al., 2005; Genge et al., 2008; Vondrak et al., 2008; Wittke et al., 2013; Folco et al., 2015; Rudraswami et al., 2015; Genge et al., 2017). The spherules range in diameter from a minimum recoverable size of 0.25 to 1.7 mm. They have been classified as I, S and G-types.

Marvin and Einaudi (1967) and others (Folco and Cordier 2015; Blanchard et al. 1980; Brownlee et al. 1997) have listed possible terrestrial and extraterrestrial origins for magnetic spherules like those discussed in this paper. Terrestrial spherules can be produced through the following mechanisms: (i) volcanism; (ii) industrial combustion of coal, crude oil, or wood; (iii) smelting products; and (iv) other phenomena such as forest fires and lightning discharges. Extraterrestrial origins may be: (i) ablation of meteorites during flight in the atmosphere (Genge et al. 1999; 2023) (ii) airbursts, i.e., disintegration of carbonaceous chondrites in flight through the atmosphere; (Van Ginneken et al. 2010; 2021; 2024) (iii) vaporization of large meteorites during impact crater formation; and (iv) infall of extraterrestrial dust particles containing iron or iron oxide, of either asteroid or comet origin (entering the atmosphere as spherules or becoming spherules in the atmosphere) (Genge et al. 2017).

On 8 January 2014 US government satellite sensors detected three atmospheric detonations in rapid succession about 84 km north of Manus Island, outside the territorial waters of Papua New Guinea (20 km). Analysis of the trajectory suggested a possible interstellar origin of the causative object CNEOS 2014-01-08: an arrival velocity relative to Earth exceeding ~45 km $s^{-1}$ (greater than the solar escape velocity) and a vector tracked back to outside the plane of the ecliptic (Siraj and Loeb 2022a). Further, the fireball light curve as measured by satellites, which showed three flares separated by a tenth of a second from each other. The bolide broke apart at an unusually low altitude of ~17 km. The object was likely substantially stronger than any of the other 272 objects in the CNEOS catalog, including the ~5%-fraction of iron meteorites from the solar system (Siraj and Loeb 2022b). Calculations of the fireball light energy suggest that about 500 kg out of an unknown total mass was ablated by the fireball and converted into ablation spherules. The fireball was localized primarily based on satellite detection of its light (Loeb 2024). The location box was also checked against a possible delay in arrival time of the direct and reflected sound waves to a seismometer located on Manus Island (Siraj and Loeb 2023), but this inference is secondary and has been debated (e.g., Brown and Borovicka 2023). In this paper, we characterize spherules retrieved in an expedition that surveyed the region identified by the US Government satellites as the meteor arrival site.

## Sampling Expedition

The expedition was mounted from Port Moresby, Papua New Guinea (PNG), to search for remnants of the bolide, labeled hereafter IM1. It utilized a 40-meter catamaran workboat, the M/V Silver Star. A 200-kg sled (**Figure 1**) was used with 300 neodymium magnets mounted on both of its sides and video cameras mounted on the tow-bridle. Approximately 0.06 $km^2$ were sampled in the target area **(Figure 2)**. The fine material collected on the neodymium magnets was extracted and brought in a wet slurry up to a laboratory set up on the bridge of the vessel for further examination. There, an initial wet-magnetic separation took place. Subsequently, both magnetic and non-magnetic separations were processed through sieves and dried. Spherules were handpicked with tweezers using a binocular zoom microscope. They ranged in size from 100 microns to 2 mm. We obtained a total of ~850 fragments consisting of spherules and shards by this method. We refer to the samples in the study as "spherules" in a descriptive way rather than to imply a particular origin.

## Spherule Samples and Location

The retrieved spherules overlap with the strewn field of the IM1 bolide based on estimates outlined in this section. This section describes our effort to retrieve material from IM1. That said, we cannot prove that any of the recovered material can be directly attributed to IM1.

The total mass of IM1 is unknown. A lower limit of 500 kg is based on the radiation energy detected from the fireball by the U.S. government satellites (Siraj & Loeb 2022a), and hence refers to just the amount of mass ablated from the bolide. According to Figure 4 of Taylor et al. (1998), the Earth-averaged flux of cosmic spherules with a diameter of 0.5–1.3 millimeters considered here, is $1.5 \times 10^{-3}/m^2/yr$, corresponding to 900 spherules over a decade across $0.06\ km^2$ and suggesting a high collection efficiency of millimeter-scale cosmic spherules by our magnetic sled (**Figure 1**) in the region it surveyed. The terminology and measurements will be explained in the upcoming subsections (see also Loeb et al. 2024a,b). **Figure 2** shows the tracks of our expedition survey. Tracks 22 and 17 were located outside the Department of Defense error box for IM1's fireball, but the full extent of IM1's strewn field is unknown.

Plinian submarine volcanic eruptions generate tephra rafts that eventually sink as the vesicles caused by exsolving gas during eruption eventually saturate with water and sink long distances away according to ocean currents. Our sled videos occasionally show this kind of volcanic material in the IM1 search area, and these boulder-like components could be from virtually anywhere in the South Pacific. Effusive submarine eruptions (e.g., Hawai'i) generate a distinctive texture called pillow lavas that we do not see in our sled videos. Subaerial volcanic eruptions can distribute ash long distances from an eruption source via prevailing winds, but these tend to be distinctive glass fragments under a microscope – not spherules. The nearest historically active volcanoes are Lou Island (~150 km south of our IM1 search zone), Manam (~450 km distant), and Rabaul (~650 km distant). Our recovered spherules are from the top centimeter of the abyssal seafloor – anything prehistoric would be below the sampling depth of our sled.

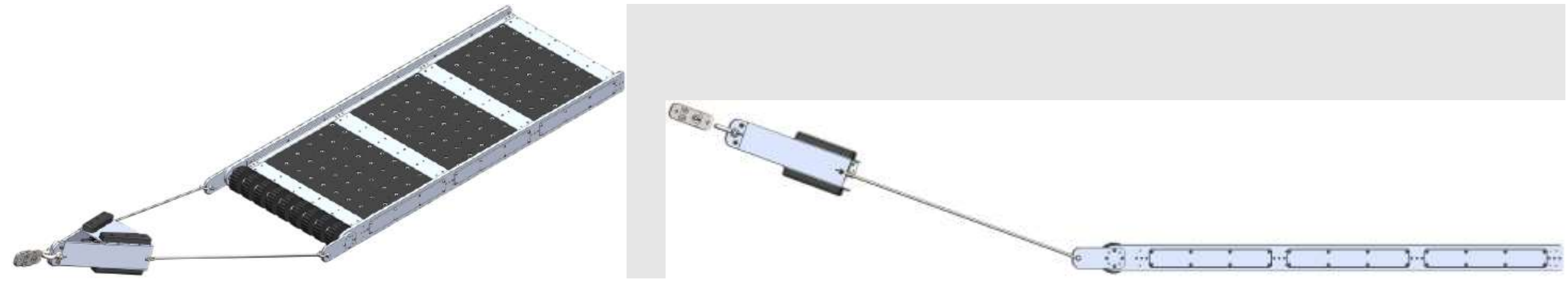

**Figure 1**: Magnetic sled design. Left: Isometric view of the magnetic sled design, Right: Side view of the sled when placed on the ocean floor. The 250-kg, 1-by-2-meter sled was covered with an array of 300 neodymium magnets on both sides and equipped with video cameras in the metal tow-halter ahead of it, which was anchored by a synthetic cable to a winch on the ship, the M/V Silver Star.

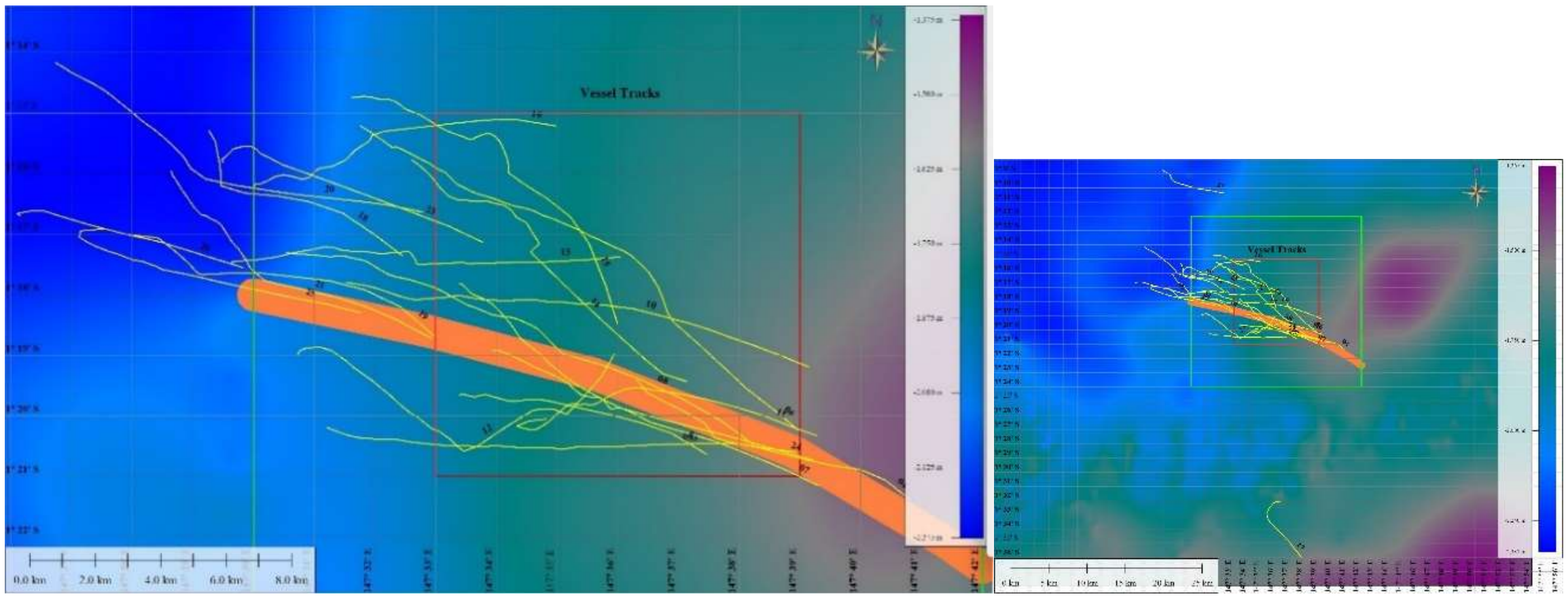

**Figure 2**: List of ship track numbers around the expected path of IM1 based on seismometer data (Siraj and Loeb, 2023) from Manus Island (orange strip). Left: Tracks within the Department of Defense error box (in red). Right: Full map including runs 22 and 17. Background colors indicate ocean depth (with scale on the right). Latitude and longitude are marked in degrees and decimal minutes. At one degree south latitude, one minute of latitude or longitude equals one nautical mile which is 1.852 km. The red box, measuring 11.112 km on a side, marks the uncertainty in the Department of Defense (DoD) localization of IM1's fireball, and the green box marks twice that size.

**Analytical Methods**

Most samples (~802) were first analyzed by microXRF with a Bruker Tornado M4 for their bulk major element composition, followed by imaging (SEM and EDS chemical mapping) and spot chemical analyses of ~80 samples with a JEOL Model JXA 8230 Electron Probe Microanalyzer (EPMA). Measurements of elemental abundances for about 60 major and trace elements were performed for 68 samples with an iCAP TQ triple quadrupole ICP-MS (ThermoFisher Scientific).

***microXRF***: The major and trace element compositions of separated spherule and shard samples was analyzed using the M4 TORNADO PLUS micro XRF instrument from Bruker, situated at the central application facility in Berlin, Germany. This instrument features a 30 W micro-focus X-ray tube and a rhodium anode, with X-rays focused through a polycapillary lens to achieve a spot size of 20 µm for the high-energy range above 20 keV. The measurements were conducted under specific conditions: 50 kV, 200 µA, and 2 mbar for a duration of 60 s. The pressure of 2 mbar was chosen to enhance the detection of light elements and minimize absorption caused by air between the sample and detector. Fluorescent signals were collected using two light element window silicon drift detectors, each with a 60 mm² active area, with independently operating signal processing units, allowing a maximal throughput of 250 kcps each at a spectroscopic resolution lower than 145 eV for manganese Kα. The samples were analyzed using the single-spot approach, capitalizing on the benefits arising from the significant information depth of X-rays (e.g. Beckhoff et al., 2006) and the beam divergence of the poly-capillary lenses (approximately 60 µm/mm). This combination ensured optimal coverage of the sample volume and yielded representative results. Sample quantification was attained through standard-less fundamental parameter quantification after instrument calibration, involving a mix of pure elements and the NIST 620 certified reference glass sample to describe lens transmission and element sensitivity. The quantification algorithm uses the forward calculation of the complete spectrum following the approach describe by Sherman (1955). To eliminate any potential for blank signals, spherules and shards were mounted on acrylic plates using double-sided tape. The results for major and trace elements were computed, presenting major elements as oxides and traces as individual elements. The primary advantage of micro-XRF lies in the combination of speed, major and trace element sensitivity as well as its non-invasiveness, positioning it as a pre-screening tool for more time-consuming and invasive analytical procedures with uncertainties below tens of percent.

***EPMA***: We used the JEOL JXA-8230 electron microprobe at the Harvard Electron Microprobe Laboratory to obtain high-resolution backscatter (BSE) and secondary (SEI) electron images, elemental X-ray maps, and chemical analyses, using focused beam of ~1 mm in diameter. Most objects were mounted on sticky tape, while several spherules were mounted in epoxy and polished. All samples were carbon-coated. The chemical compositions of intact objects were measured by energy dispersive spectroscopy (EDS) using factory calibration curves. The polished areas of several spherules were analyzed by wavelength dispersive spectroscopy (WDS) using common natural minerals and synthetic glasses as calibration standards (e.g., Petaev and Jacobsen, 2009). The EDS analyses and imaging we performed at accelerating voltage of 20 kV and beam currents of 5–10 nA and ~0.1 nA at low- (< 1000×) and high-resolution (> 2000×), respectively. In the WDS analyses we used accelerating voltage of 15 kV, beam current of 20 nA, and counting times of 30 sec and 15 sec on peak and background, respectively.

***TQ-ICP-MS*** (Triple Quad elemental analysis): Measurements of elemental abundances for major and trace elements were performed on the iCAP TQ quadrupole ICP-MS (ThermoFisher Scientific) in the Cosmochemistry Laboratory at Harvard University. USGS reference materials were thoroughly dissolved and diluted in a 2% $HNO_3$ solution spiked with 10 ppb indium diluted to a factor of 5000 to be used as standards. Spherules were prepared for mass spectrometry measurements by first individually digesting the samples in a mixture of concentrated HF-$HNO_3$-HCl at a 1:3:1 ratio at 120–140 ºC overnight. The samples were subsequently dried down and then redissolved in a second acid mixture involving an aqua regia solution mixed with $H_2O$ at a 3:2 ratio and heated to 120-140 ºC overnight. This dissolution was dried down for the second time and redissolved in a high-purity 2% $HNO_3$ solution. A small aliquot (3%) was drawn from this solution and further diluted for elemental analysis. To account for and to correct instrumental drift, the 2% $HNO_3$ solution used for dilution was spiked with 2 ppb indium as an internal standard, prepared identically to the standard solutions. Measurements were performed in KED mode with He as a collision cell gas as recommended by the Reaction Finder function built in the iCAP TQ Qtegra software, except for Cr, which was measured in TQ mode with $O_2$ as a mass-shifted molecule. The prepared spherule solutions were measured as an unknown against a four-point calibration line consisting of a blank and three USGS standards: BCR-2, BHVO-2, and AGV-2. Calibration curves for individual elements were checked for linear intensity to concentration correlations for accurate measurements. Routine measurements of AGV-2 as an unknown on the iCAP TQ suggest fractional errors to be within 6% using this method.

## Results

### *Imaging and morphology of the spherules*

Electron microprobe images of recovered spherules (and in some cases partially melted micrometeorites) are shown in **Figures. 3, 4 and 5.** The spherules in **Figure 3** are what we call primitive spherules, identified as such by their bulk chemical composition (see next section). The dendritic textures of these spherules suggest rapid cooling. The spherules/micrometeorites in **Figure 4** and **5** are all identified as differentiated spherules by their chemical compositions (see next section) and do not in general have as perfect a spherical shape as the primitive group, and some of them are clusters of coalesced spherules. The complete spherule dataset will be released in a future publication. In addition to electron microprobe images, low resolution photos were taken for 684 samples prior to micro-XRF analysis. Among these, ~64 were determined to be "shards" as opposed to "spherules," which were discerned based on sharp angles resembling broken surfaces or evidence of fragmentation.

An example of a large (1.3 mm in maximum diameter, 0.9 mm average) spherule in the region along IM1's likely path is S21 (IS14-SPH1) from run 14. This lopsided spherule, shown in **Figure S4**, is a composite of three spherules that solidified shortly after merger but too late for the merger product to become spherical. The mass of S21 (1.7 mg) is about twice that of IS16A SPH1 (0.84 mg). The existence of a triple-merger like S21 can potentially be explained as a product of a meteoric airburst (Genge et al (1999); Van Ginneken et al., 2012, 2021, 2024). The total mass collected by S21-like spherules in all our runs is of order ~0.1g. Given the sled's width of 1m, the total surveyed area, ~0.06 $km^2$, constitutes a fraction of $\sim 10^{-3}$ of IM1's strewn field. This implies a total mass in S21-like spherules in our estimated strewn field of order ~100g, as expected given that most of IM1's mass evaporated to undetectable particles (well below tens of µm in size) or gas (Tillinghast-Raby et al., 2022). Assigning a mass of $\sim 10^{-3}$ g per spherule implies a total number of $\sim 10^5$ such spherules. Based on IM1's speed and fireball energy, the total mass ablated by IM1's fireball is $\sim 5 \times 10^5$ g corresponding to an object radius of R ~50 cm (Siraj and Loeb, 2022b). The total number of spherules divided by the initial volume associated with R yields an initial spherule number density of $n \sim 0.2\ cm^{-3}$ which gets diluted as the material expands. For the characteristic diameter of a spherule, ~1mm, the geometric cross-section for spherule-spherule collisions is $\sigma \sim 4 \times 10^{-2}$ $cm^2$. The resulting collision probability is $\tau \sim n\sigma(2R) \sim (0.2\ cm^{-3}) \times (4 \times 10^{-2}\ cm^2) \times (100\ cm) \sim 0.8$, implying a likelihood of $\tau^2 \sim 0.6$ for triple-spherule mergers such as S21. Mergers that occur inside a liquid envelope would result in a spherical shell with embedded sub-spherules inside of it. Mergers are also likely amongst volcanic dust.

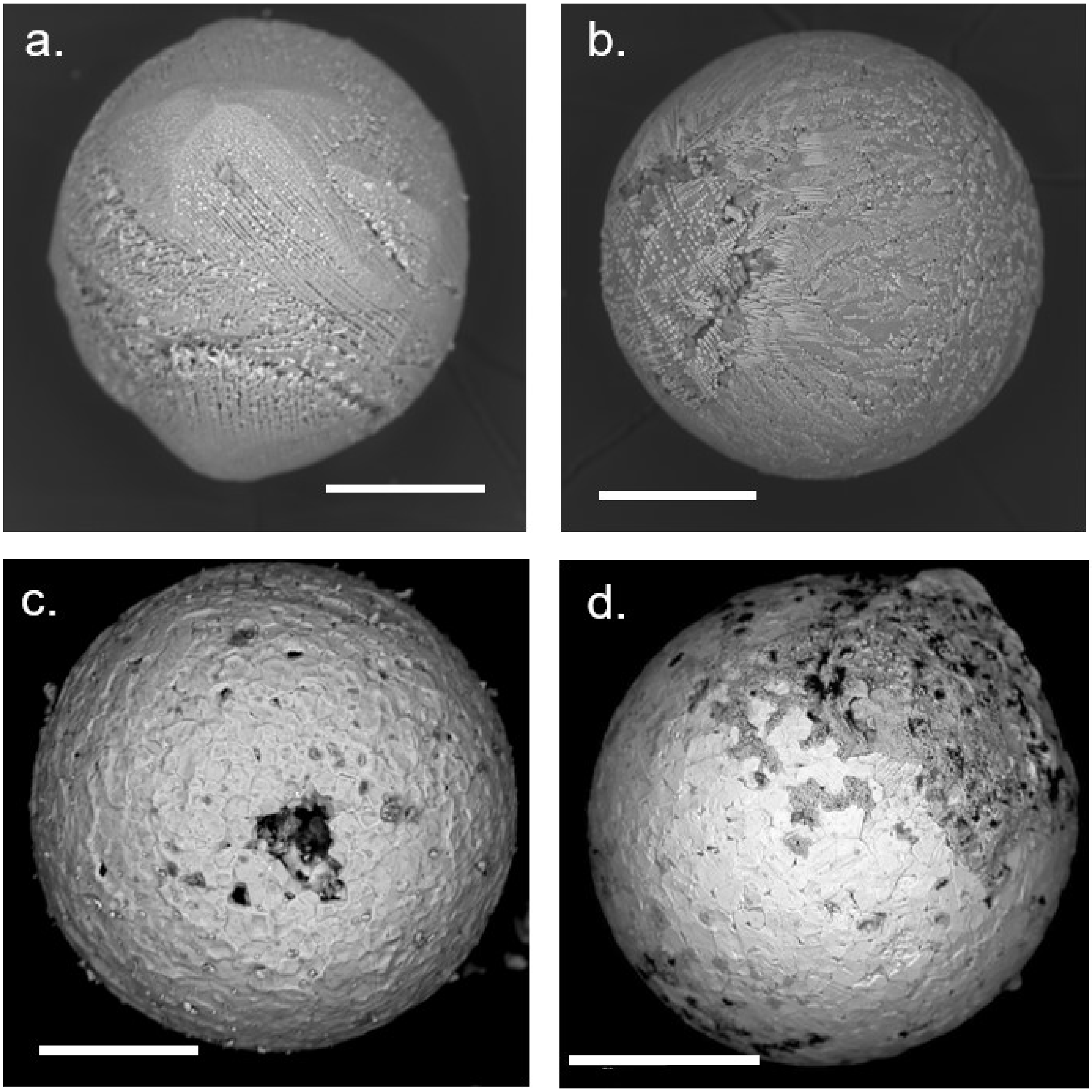

**Figure 3.** BSE images of primitive spherules from the a) S-type (IS20M-1) b) G-type (IS20M-21) c) I-type, high Ni (IS8M2-20) and d) I-type, low Ni (IS16A-SPH1) groups. The scale bar is 100 microns.

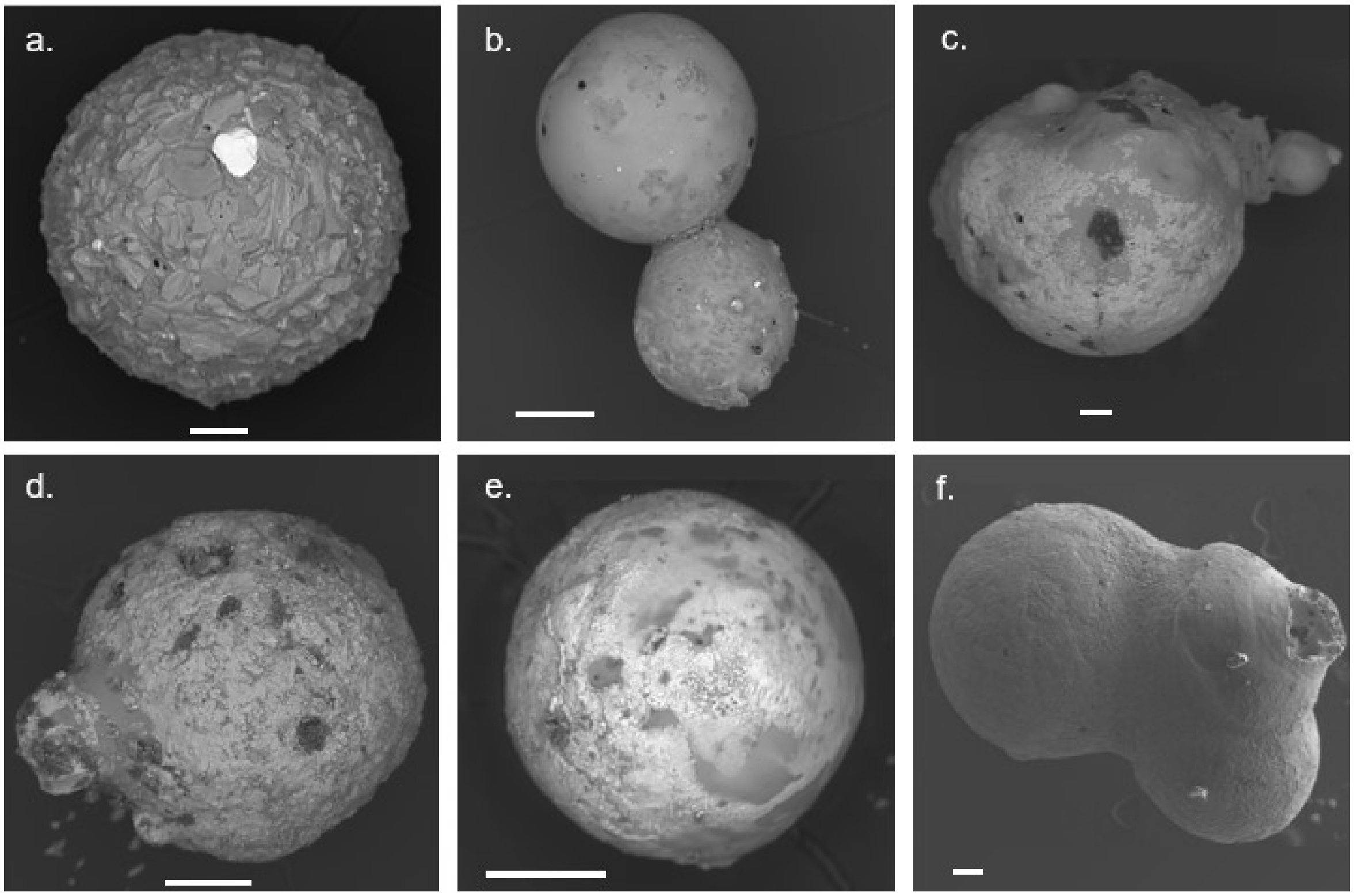

**Figure 4**. BSE images of differentiated spherules from each of the differentiated (D-type) spherule classifications: a) D-type low Sr, high Si (IS11M2-2; porphyritic) b) D-type high Sr, high Si (IS19M-14) c) D-type, high Si (17NMAG-5; vitric) d) D-type, low Sr, low Si (IS21-4; cryptocrystalline) e) D-type high Sr, low Si (IS14M-2; cryptocrystalline) and f) D-type, low Si (S21 or IS14-SPH1; vitric) groups. The images c) and f) are also classified as "BeLaU"-type spherules. The scale bar is 100 microns. See the text for the criteria for the different classifications of the spherules.

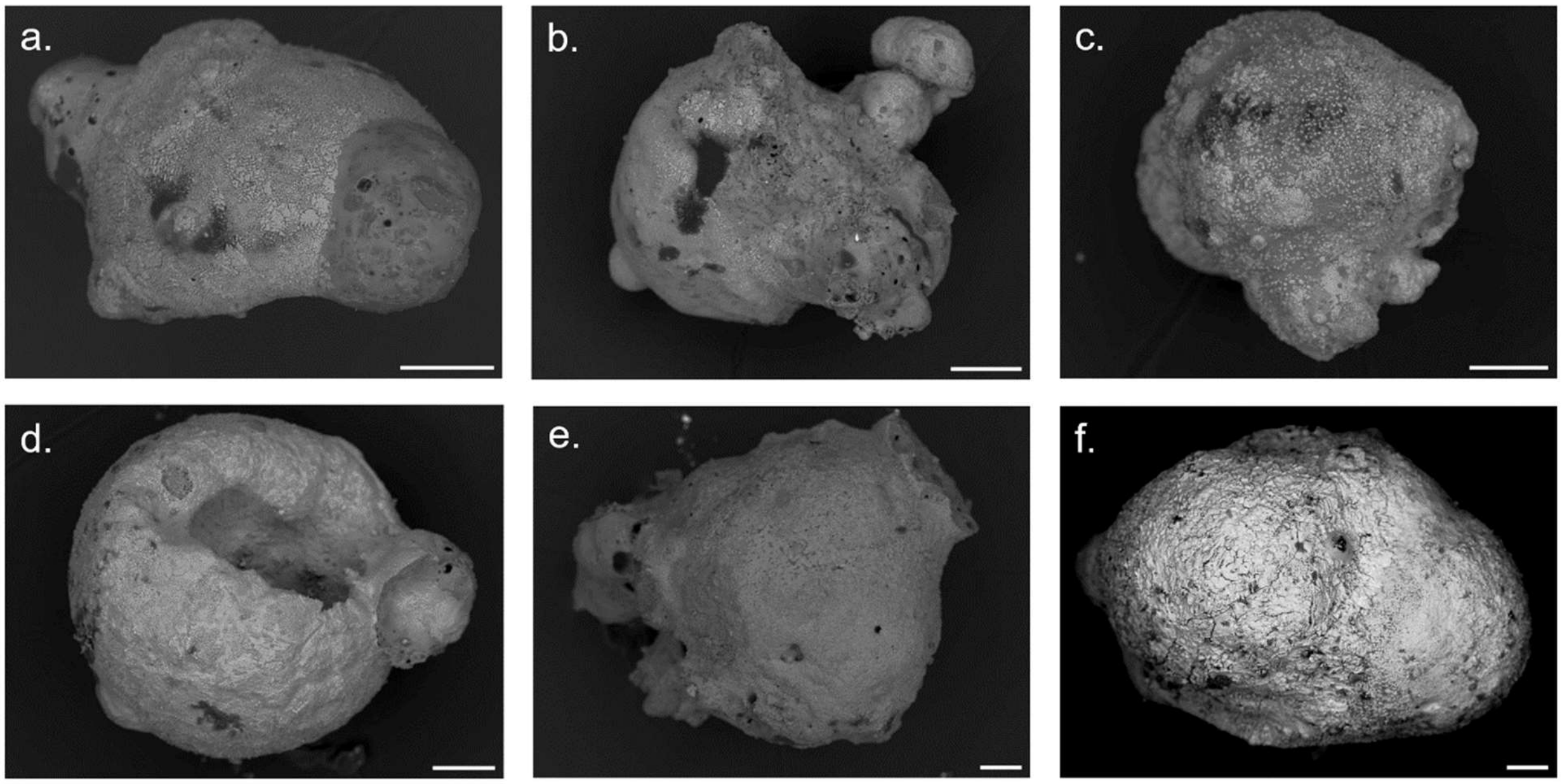

**Figure 5**. Additional BSE images of D-type scoriaceous (partially melted) micrometeorites. : a) 19NMAG-52 (D-type, high Sr, low *Si/(Mg+Fe+Si)*), b) 19MAG-50 (D-type, high Sr, low *Si/(Mg+Fe+Si)* ), c)

19MAGx-4 (BeLaU, low *Si/(Mg+Fe+Si)* ), d) 17MAG-2 (BeLaU, low *Si/(Mg+Fe+Si)* ), e) 17MAG-29 (BeLaU, low *Si/(Mg+Fe+Si)* ), and f) 17MAG-6 (D-type, high Sr, low *Si/(Mg+Fe+Si)* ). The scale bar is 100 microns. See the text for the criteria for the different classifications of the spherules.

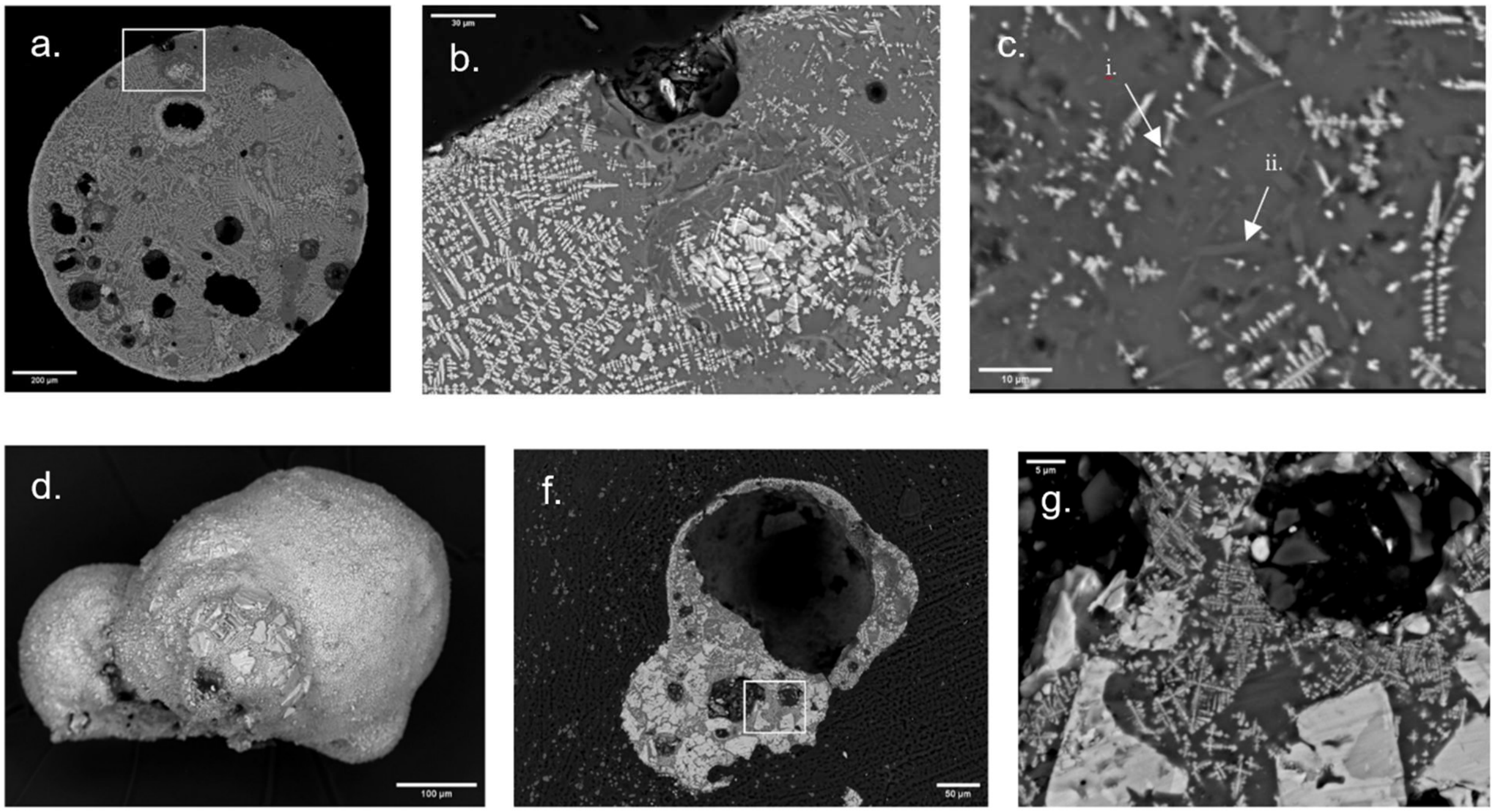

**Figure 6**. a) Cross section of IS14 SPH4, determined to be a BeLaU-type spherule. The spherule is vesicular and exhibits dendritic textures and a dendritic magnetite/spinel rim. b) Close-up of the cross-section for IS14 SPH4, exhibiting dendritic crystals in a silicate mesostasis, a concave vesicle (center-right) and a dendritic magnetite rim. c) Another close-up image of IS14 SPH4 with i) dendritic spinels and magnetites and ii) aluminosilicate crystals. d) 17NMAG-28, a high Sr, low Si D-type spherule that is rich in Fe. e) Cross section of 17NMAG-28, exhibiting vesicles and a large concavity, relict grains composed of Fe-Al-oxide and what appears to be a partial magnetite/spinel rim. f) Close-up of the cross-section for 17NMAG-28, showing dendritic patterns, and relict grains in a silicate mesostasis. Scale bars: a) 200 microns, b) 30 microns, c) 10 microns, d) 100 microns, e) 50 microns, and f) 5 microns. See the text for the criteria for the different classifications of the spherules.

**Analysis of Polished Sections**

Cross-section backscattered electron images of two highly differentiated spherules (IS14 SPH4 and 17NMAG-28), further discussed in the next sections, are shown in **Figure 6**. The two samples are both characterized by vesicles and dendritic patterns composed of magnetite and spinel in silicate mesostasis with alumnosilicate crystals throughout their cross-section images. IS14 SPH4 (**Figure 6a, b, c**) exhibits a magnetite rim with dendritic patterns (**Figure 6b**), and 17NMAG-28 (**Figure 6d, e, f**) also exhibits what appears to be a spinel or magnetite rim (**Figure 6e**). Relict grains composed of Fe-Al-oxide are apparent in **Figure 6e** and **f**.

**Classification of the spherules based on elemental compositions.**

Cosmic spherules are sub-divided into three compositional types (Blanchard et al., 1980, Genge et al., 2008) based on their bulk chemistry. These are the silicate-rich spherules or S-type, the Fe-rich spherules or I-type and glassy spherules or G-types. There is strong evidence that S-type cosmic spherules with chondritic-like composition are related to carbonaceous chondrites and ordinary chondrites (Genge et al., 2008; Cordier 2014, Rudraswami et al. 2016, Van Ginneken 2017). It has been suggested that I-type spherules are metal grains released from carbonaceous chondrites disaggregated in space (Herzog et al. 1999). The glassy or G-type spherules are thought to be mixtures of S and I. Relatively rare spherules have been called *differentiated* as they have similarities to achondrite meteorites and have been treated as a subgroup of S-type spherules. Differentiated spherules have major-element compositions with higher Si/Mg and Al/Si ratios, and higher refractory lithophile trace element contents relative to chondritic spherules (e.g., Folco and Cordier 2015).

The bulk major element, Sr and Ni compositions of 745 spherules from the IM1 site measured by micro-XRF are in **Table S1** (supplement). The data are plotted in a Mg/Si histogram (**Figure 7**) and show a minimum at Mg/Si = 1/3. Spherules with Mg/Si > 1/3 are similar to chondritic meteorites and cosmic S-type spherules, while those with Mg/Si <1/3 are similar to igneous rocks from Earth and other planetary bodies. The data are also plotted in a Mg-Si-Fe ternary diagram (**Figure 8**), since such a diagram has been shown to effectively distinguish the S-, I- and G-type groups (e.g., Folco and Cordier 2015). We note that whole rock compositions were used as comparisons in this figure. We also note that there are two distinct groups of spherule compositions in this plot. About 78% of the spherules fall along the trend of S-, G- and I-type spherules. These are referred to as primitive spherules as they are thought to be related to primitive chondritic meteorites and represent materials that have not gone through planetary differentiation. The remaining 22% of the spherules have low Mg and plot close to the Si-Fe side of the diagram. The high-Si part of this group plots within the range of terrestrial igneous rocks that are shown for comparison. These spherules are thus called differentiated, meaning they are likely derived from crustal rocks of a differentiated planet. Since they are clearly different from the differentiated subgroup of S-type spherules we give them a new name *D-type spherules*. The term "differentiated" is typically attributed to spherules with achondrite origins of the HED class of meteorites (e.g., Cordier et al. 2011b, Folco and Cordier, 2015). However, as these D-type spherules are substantially different from those in that they have higher incompatible element abundances as well as more fractionated incompatible element patterns, we attribute a new classification to emphasize their distinction. The Mg/Si = 1/3 is used to distinguish primitive and differentiated spherules. The D-type spherules were discovered in 17 of the 24 tracks explored during this expedition.

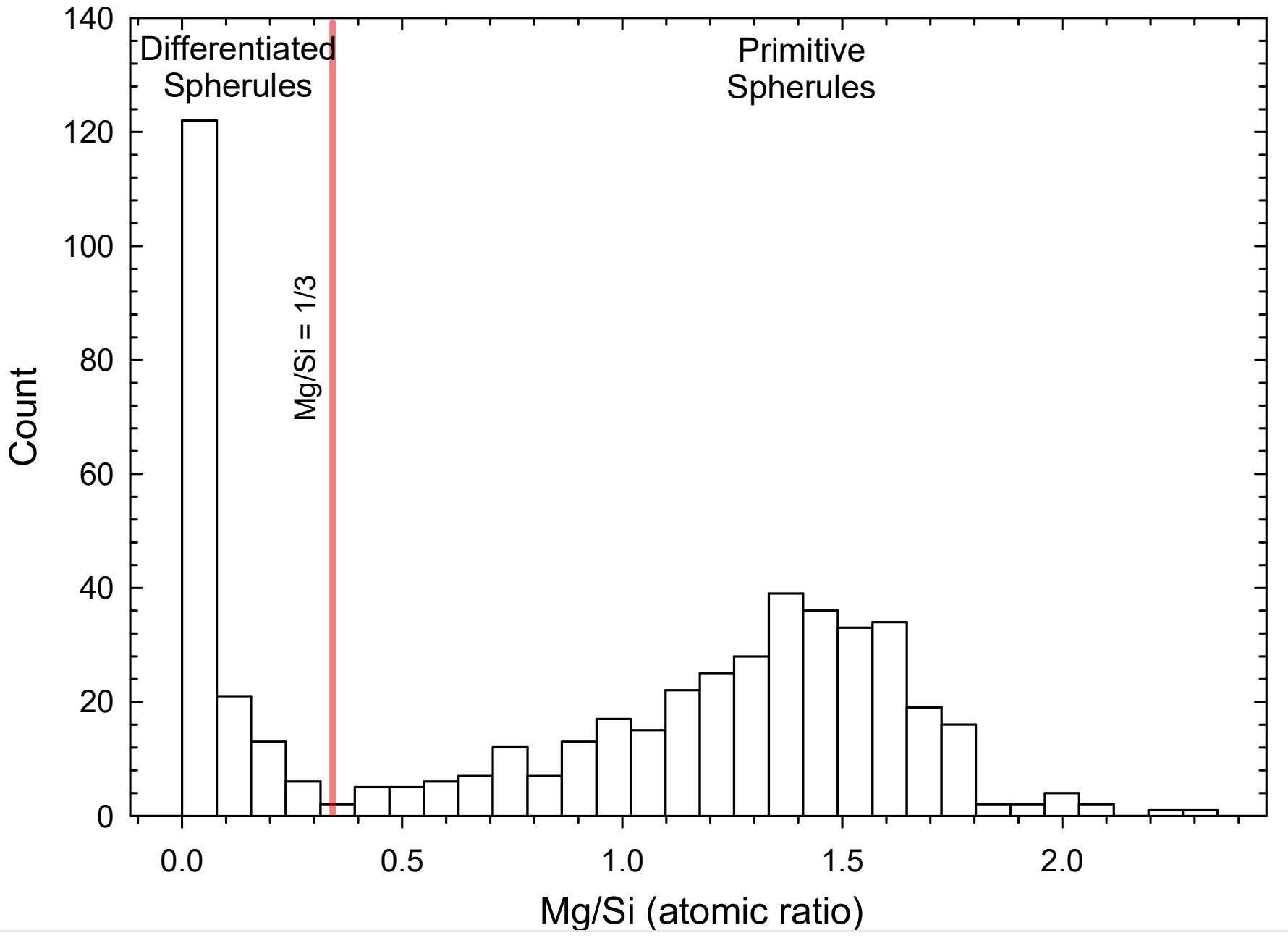

**Figure 7**. Histogram of the Mg/Si ratio measured with micro-XRF for 745 IM1 site spherules. This diagram shows a clear dividing line between primitive and differentiated spherules.

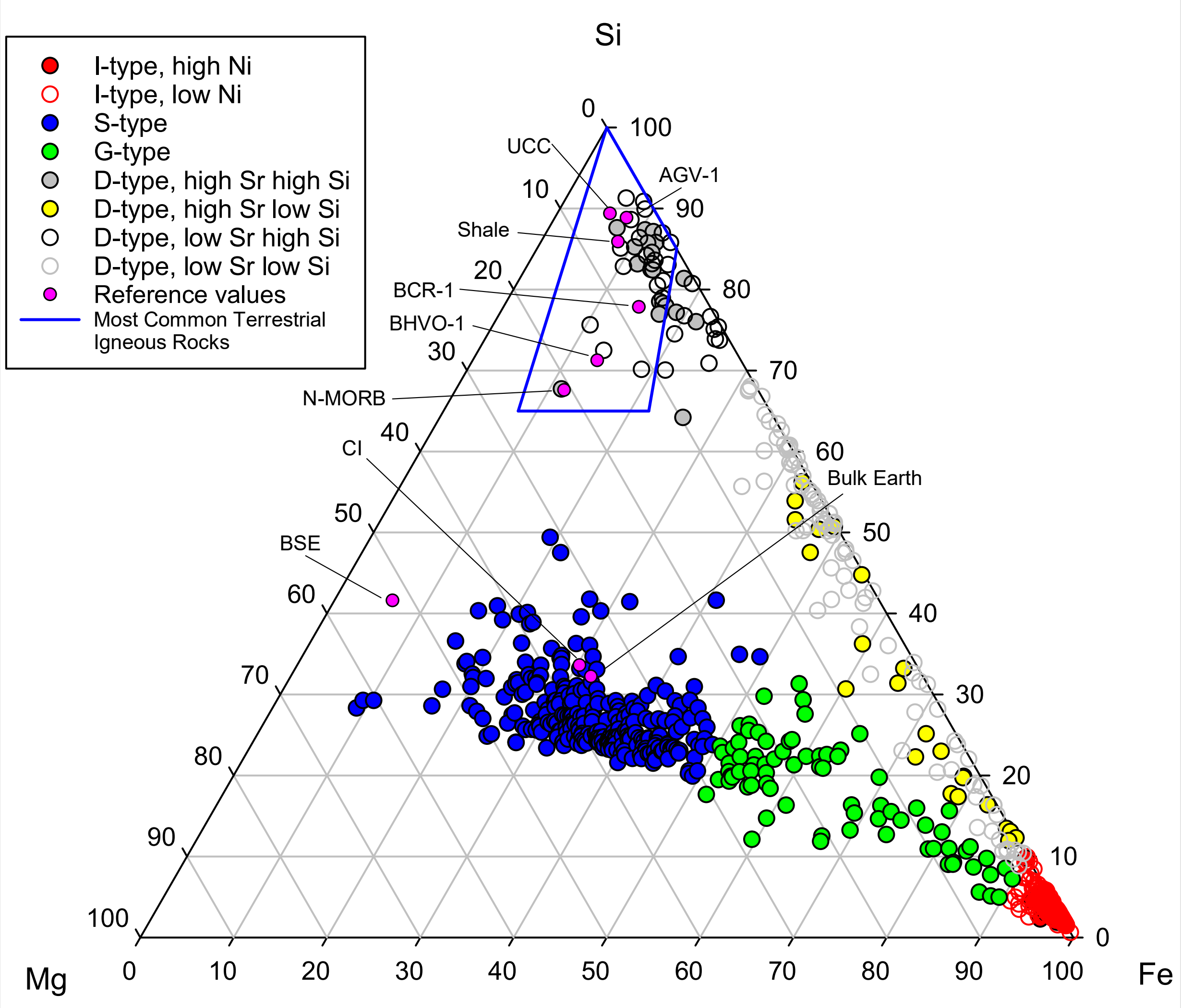

**Figure 8**. Atomic Mg-Si-Fe plot (atom %) of micro-XRF bulk compositional data for 745 IM1 site spherules and shards. The spherules are classified with the parameter values in **Table 1**. The spherule groups are compared to reference values of Earth materials (Bulk Earth, bulk silicate Earth [BSE], upper continental crust [UCC], shale, normal midocean ridge basalts [N-MORB], Hawaiian basalt [BHVO-1], Columbia River basalt [BCR-1], Guano Valley andesite [AGV-1]) and CI meteorites. Also shown is the range of chemical compositions of terrestrial igneous rocks.

For the primitive spherules we use 100Fe/(Fe+Si+Mg) > 90 to distinguish I-types from S- and G-types. This group is supposed to be primarily made of iron compounds and is clearly distinct from the other primitive spherules in the 100Fe/(Fe+Si+Mg) histogram (**Figure 9a**). This I-group is further subdivided into high Ni (>4000 ppm) and low Ni (<4000 ppm) groups (**Figure 9b**), as high Ni spherules are most likely of cosmic origin and typically have Ni > 4000 ppm (Engrand et al. 2006). The S- and G-type dividing line is 100Si/(Fe+Si+Mg) = 50 (**Figure 9a**), relatively consistent with previous literature (e.g., Brownlee et al. 1997; Taylor et al. 2000; Folco and Cordier 2015). The primitive spherule groups are compared to reference values of Bulk Earth, bulk silicate Earth (BSE) (McDonough and Sun 1995), and CI meteorites (Anders and Grevesse 1989) that are and should be on the primitive trend (**Figure 8**).

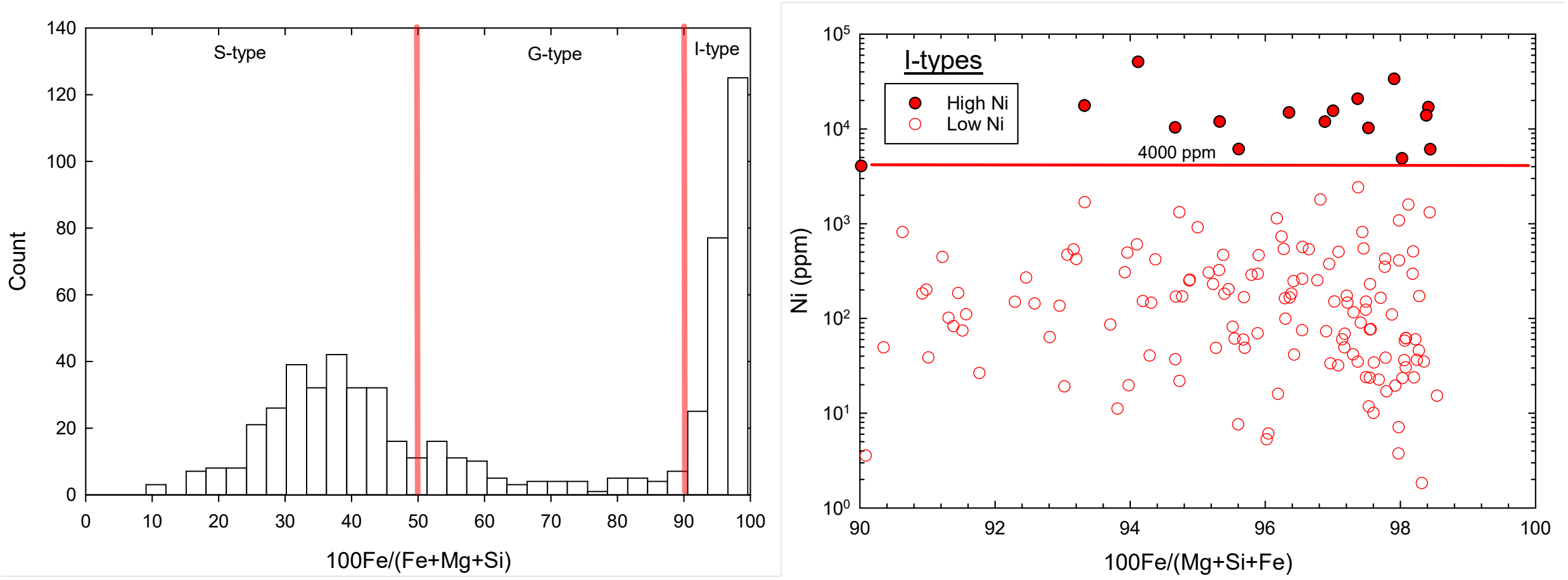

**Figure 9**. Left: Histogram of the 100Fe/(Fe+Si+Mg) measured with micro-XRF for 745 IM1 site spherules. This diagram shows the dividing lines selected between S-, G-, and I-types. Right: Plot of Ni(ppm) vs. 100Fe/(Fe+Si+Mg), showing the dividing line selected for high-Ni vs low-Ni spherules.

Owing to the incompatible nature of Sr, the Sr content of differentiated spherules is a good indication of the enrichment of refractory lithophile elements in the spherules and thus the extent of differentiation. The differentiated spherules in this study are thus divided into high Sr (>450 ppm) and low Sr groups (<450 ppm). They are further subdivided on the Si-content. For high Sr spherules we use 100Si/(Fe+Si+Mg) = 60 as the dividing line and for low Sr spherules a value of 100Si/(Fe+Si+Mg) = 70. These dividing lines are shown in **Figure 10**. The spherule groups are compared to reference values for samples of the Earth's crust in **Figure 8**. This includes average composition of normal midocean ridge basalts (N-MORB) (Gale et al. 2013), average upper continental crust (UCC) (Rudnick and Gao 2014) and shale (Ray and Paul 2021). Also plotted is the outline of the field of 37000 terrestrial igneous rocks as well as 3 USGS standards [Hawaiian basalt (BHVO-1), Columbia River basalt (BCR-1), Guano Valley andesite (AGV-1) (Jochum et al. 2007)].

We note that the high Si varieties of D-spherules plot close to or within the range of terrestrial igneous rocks, while the low Si groups do not. Thus, the D-type spherules and have been divided into four distinct groups based on their bulk composition. This results in eight distinct spherule groups that are all shown in **Figure 8**. The parameters used to subdivide the spherules are summarized in **Table 1.** The total number of spherules identified in each group are listed in **Table 2**.

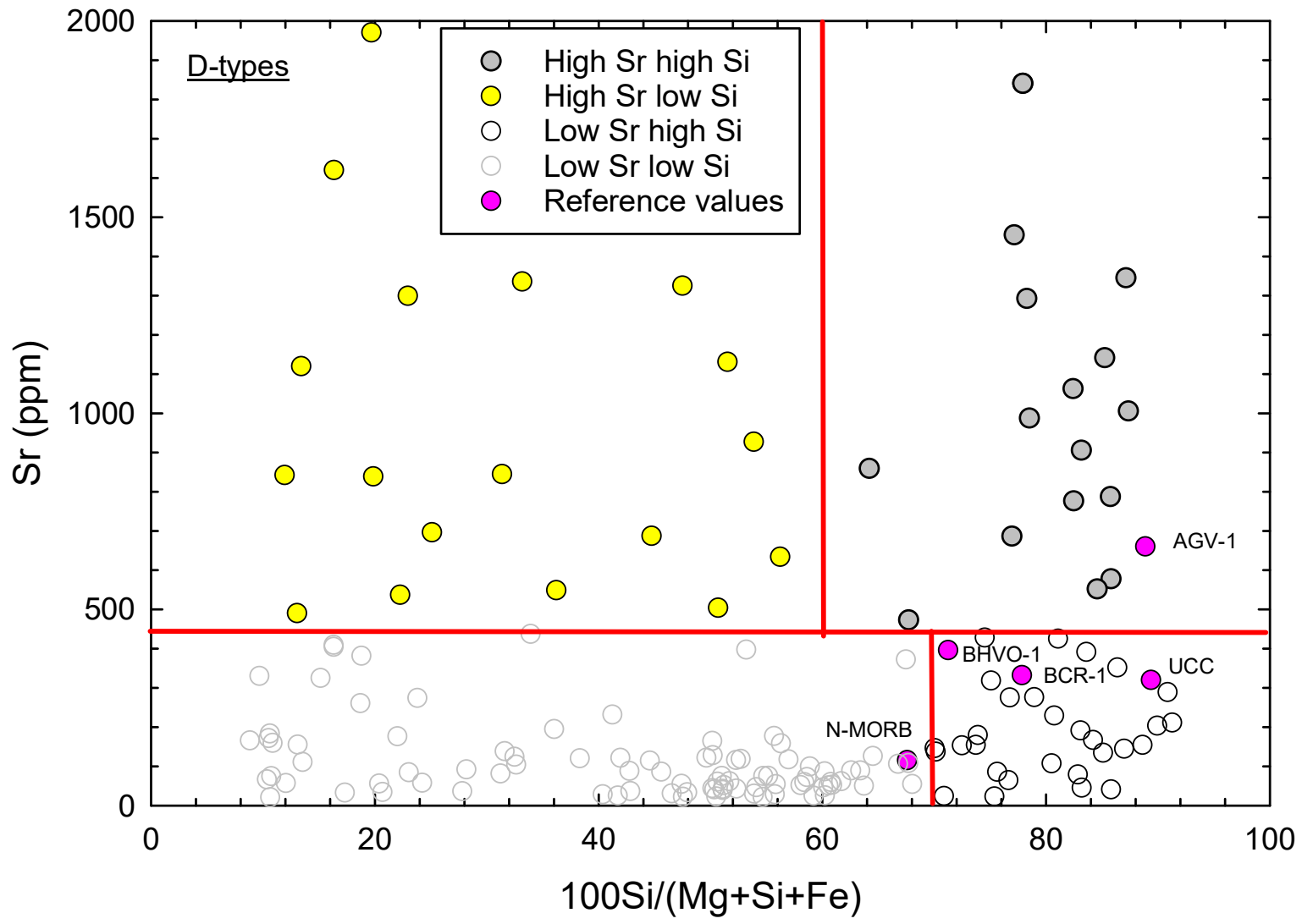

**Figure 10**. Micro-XRF bulk compositional data for 745 IM1 site spherules. Plot of Sr(ppm) vs. 100Si/(Fe+Si+Mg).

**Table 1**. Classification of the spherules with the Mg-Si-Fe diagram.

| | Type | 100Fe/ (Fe+Si+Mg) | Ni (ppm) | Mg/Si | Sr (ppm) | 100Si/ (Fe+Si+Mg) |
|---|---|---|---|---|---|---|
| Primitive spherules | I-type, high Ni | >90 | >4000 | | | |
| | I-type, low Ni | >90 | <4000 | | | |
| | S-type | <50 | | >1/3 | | |
| | G-type | >50 and <90 | | >1/3 | | |
| Differentiated spherules | D-type, high Sr, high Si | | | <1/3 | >450 | >60 |
| | D-type, high Sr, low Si | | | <1/3 | >450 | >60 |
| | D-type, low Sr, high Si | | | <1/3 | <450 | >70 |
| | D-type, low Sr, low Si | | | <1/3 | <450 | >70 |

**Table 2**. Comparison of the number of each spherule type identified by micro-XRF and by ICP-MS results.

| | **Type** | **Micro-XRF Number** | **Micro-XRF Percent** | **ICP-MS Number** | **ICP-MS Percent** |
|---|---|---|---|---|---|
| Primitive spherules | I-type, high Ni | 18 | 2.4 | 5 | 7.4 |
| | I-type, low Ni | 212 | 28.5 | 18 | 26.5 |
| | S-type | 275 | 36.9 | 19 | 27.9 |
| | G-type | 78 | 10.5 | 0 | 0 |
| Differentiated spherules | D-type, high Sr, high Si | 20 | 2.7 | 6 | 8.8 |
| | D-type, high Sr, low Si | 23 | 3.1 | 5 | 7.4 |
| | D-type, low Sr, high Si | 29 | 3.9 | 0 | 0 |
| | D-type, low Sr, low Si | 90 | 12.1 | 3 | 4.4 |
| | D-type, BeLaU high Si | | | 2 | 2.9 |
| | D-type, BeLaU low Si | | | 10 | 14.7 |
| | Total | 745 | 100 | 68 | 100 |

The major element and trace element bulk compositions of 68 spherules from the IM1 site measured by ICP-MS are in **Table S2** (supplement). The ICP-MS data for the spherules include many more elements than the micro-XRF bulk composition data but does not include measurement of Si. Thus, we also plot the micro-XRF data in a Mg-Al-Fe diagram (**Figure 11a**) and compare it to the same diagram with ICP-MS measurements (**Figure 11b**). The micro-XRF results show the same groups in the Mg-Al-Fe plot as for the Mg-Si-Fe plot in **Figure 8.** Note that the primitive and differentiated trends are even better separated in **Figure 9a**, and the fields of the primitive spherule groups are also well defined by this diagram. We conclude that this new diagram is a good alternative to the Mg-Si-Fe diagram, when Si data are missing. The Sr concentrations of the D-type spherules measured by micro-XRF are plotted versus the Al/Fe ratio in **Figure 12a**. While there is some overlap, this diagram gives a relatively good separation of the high Si and low Si groups after separating the differentiated spherules into high (>450 ppm) and low (<450 ppm) Sr groups and is thus also useful for data with missing Si measurements. While distinctions between "high" and "low" Si content here strictly refer to Si content normalized to Si+Fe+Mg, we refer to this as "high Si" and "low Si" groups throughout this study. Classification of the spherules with the parameters in **Table 3** will thus yield a relatively equivalent classification of the spherules compared to that obtained with the Mg-Si-Fe diagram. The primitive spherules are identified by 100Al/(Fe+Al+Mg) <10 and the differentiated spherules by 100Al/(Fe+Al+Mg) >10. For the primitive spherules the I-types are identified by 100Fe/(Fe+Al+Mg) >90, the S-types by 100Fe/(Fe+Al+Mg) <65, and the G-types in between these two values.

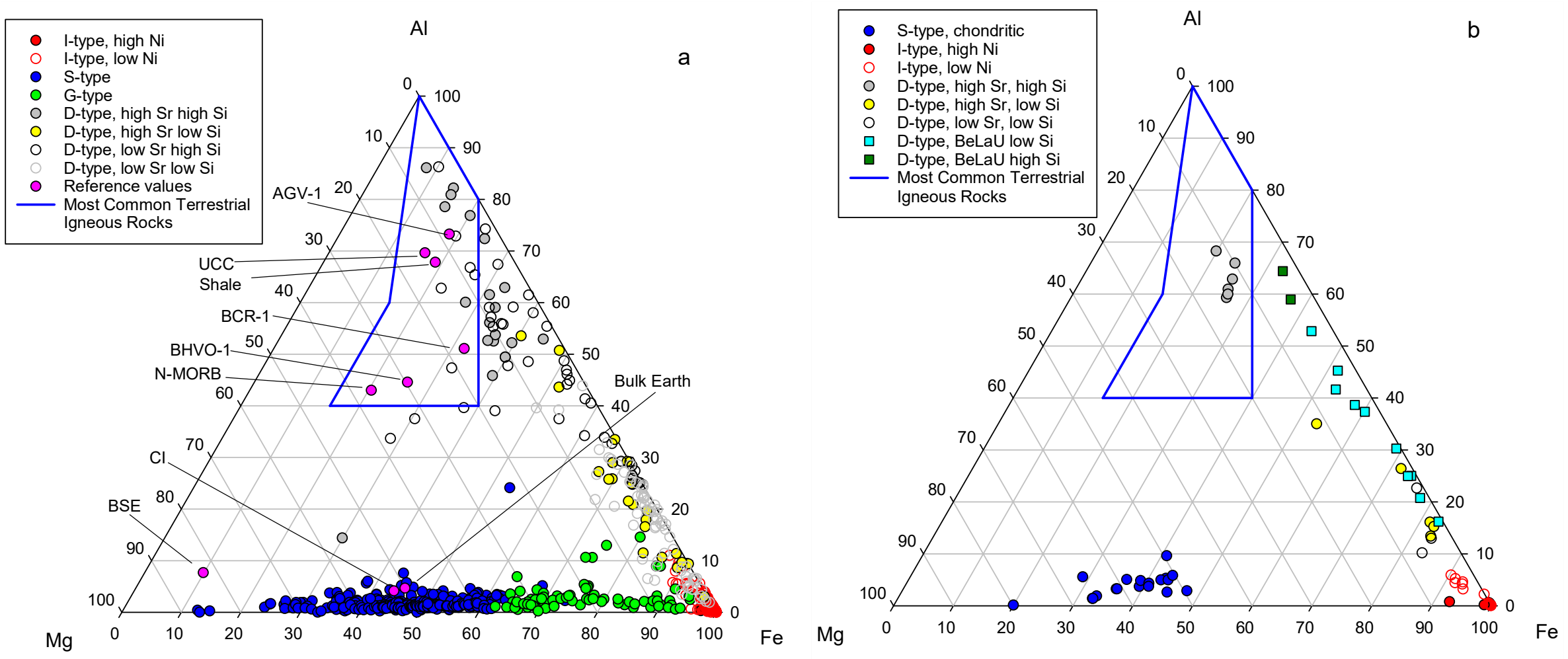

**Figure 11** a) Atomic Mg-Al-Fe plot (atom %) of micro-XRF bulk compositional data for 745 IM1 site spherules. The spherules are classified with the parameter values in **Table 1** and are the same spherules shown in **Figure 9**. b) ICP-MS data for 68 spherules, classified with the parameters in **Table 3** as discussed in the text.

**Table 3**. Classification of the spherules with the Mg-Al-Fe diagram.

| | Type | 100Fe/ (Fe+Al+Mg) | Ni (ppm) | 100Al/ (Fe+Al+Mg) | Sr (ppm) | Sr vs Al/Fe | BeLaU |
|---|---|---|---|---|---|---|---|
| Primitive spherules | I-type, high Ni | >90 | >4000 | <10 | | | |
| | I-type, low Ni | >90 | <4000 | <10 | | | |
| | S-type | <65 | | <10 | | | |
| | G-type | >65 and <90 | | <10 | | | |
| Differentiated spherules | D-type, high Sr, high Si | | | >10 | >450 | See Figure 12 | <80 |
| | D-type, high Sr, low Si | | | >10 | >450 | See Figure 12 | <80 |
| | D-type, low Sr, high Si | | | >10 | <450 | See Figure 12 | <80 |
| | D-type, low Sr, low Si | | | >10 | <450 | See Figure 12 | <80 |
| | D-type, BeLaU, high Si | | | >10 | | | >80 |
| | D-type, BeLaU, low Si | | | >10 | | | >80 |

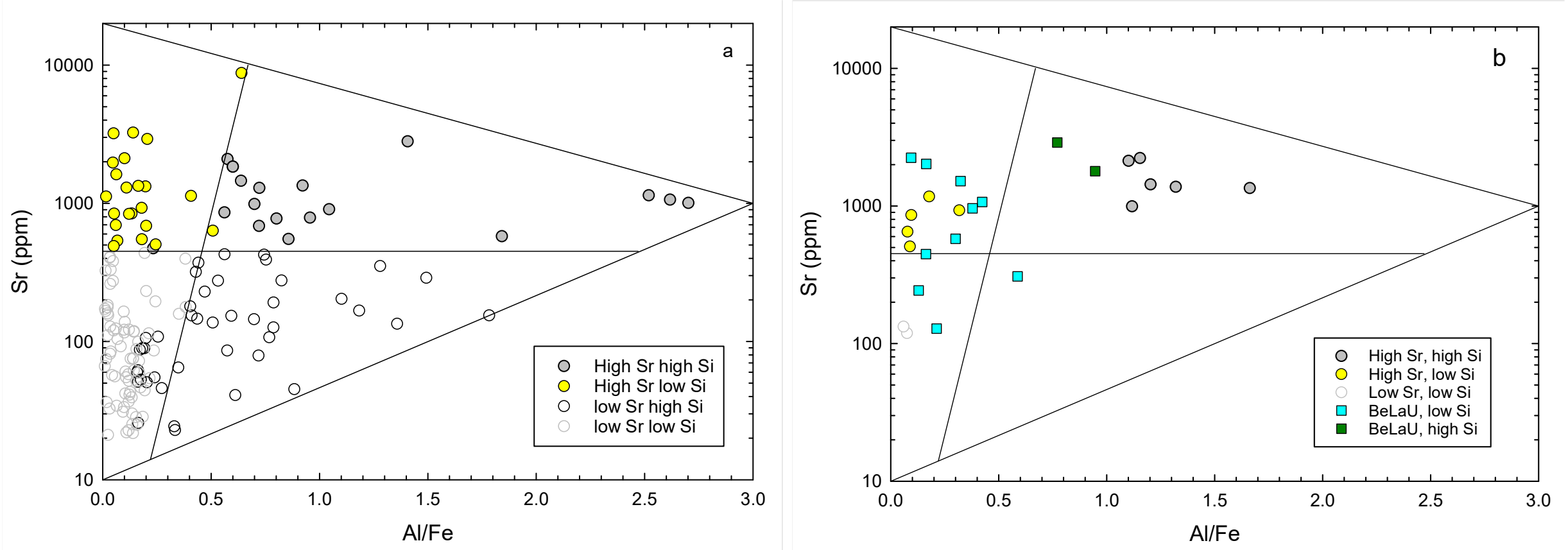

**Figure 12**. Sr (ppm) vs the Al/Fe weight ratio for D-type spherules from the IM1 site. a) Micro-XRF data, b) ICP-MS data.

For differentiated spherules we first use the fields of the four different D-types shown in **Figure 12a** to identify the same types for the ICP-MS data. In addition, we use **Figure 13** to identify spherules with particularly high contents of refractory lithophile elements, based on the enrichments of Be, La and U relative to Mg and Fe. First the concentrations of these elements are normalized to the same elements in CI chondrites. The CI-normalized values ($Be_{CI}$, $La_{CI}$, $U_{CI}$, $Mg_{CI}$ and $Fe_{CI}$) are used to calculate the following plotting parameters for the ternary diagram in **Figure 13**:

$$BeLaU = \frac{Be_{CI} + La_{CI} + U_{CI}}{Be_{CI} + La_{CI} + U_{CI} + Mg_{CI} + Fe_{CI}}$$

$$M = \frac{1000(Mg_{CI})}{Be_{CI} + La_{CI} + U_{CI} + Mg_{CI} + Fe_{CI}}$$

$$F = \frac{100(Fe_{CI})}{Be_{CI} + La_{CI} + U_{CI} + Mg_{CI} + Fe_{CI}}$$

The BeLaU parameter was divided by 100, while the M-parameter was multiplied by 10 to make use of the entire area inside the ternary diagram. We define a BeLaU-type spherule as having BeLaU > 80, and we define high and low Si varieties based on the Al/Fe ratio, as in **Figure 12a**. This procedure identifies 10 of D-type spherules as low Si BeLaU-type spherules and 2 as high Si BeLaU-type spherules as shown in **Figure 13**.

BeLaU-type spherules can only be identified with ICP measurements. We found 12 BeLaU-type spherules out 26 D-type spherules identified with ICP measurements (**Table 2**). The BeLaU spherules were found in tracks 4, 13, 14, 17 and 19. With the micro-XRF measurements, we identified 162 D-type spherules out of a total of 745 spherules. This results in an estimate of 10 % of all spherules to be of BeLaU-type. Thus, out of the ~802 spherules identified as natural materials by our micro-XRF and ICPMS measurements we estimate that there should be up to 80 BeLaU-type spherules in our collection.

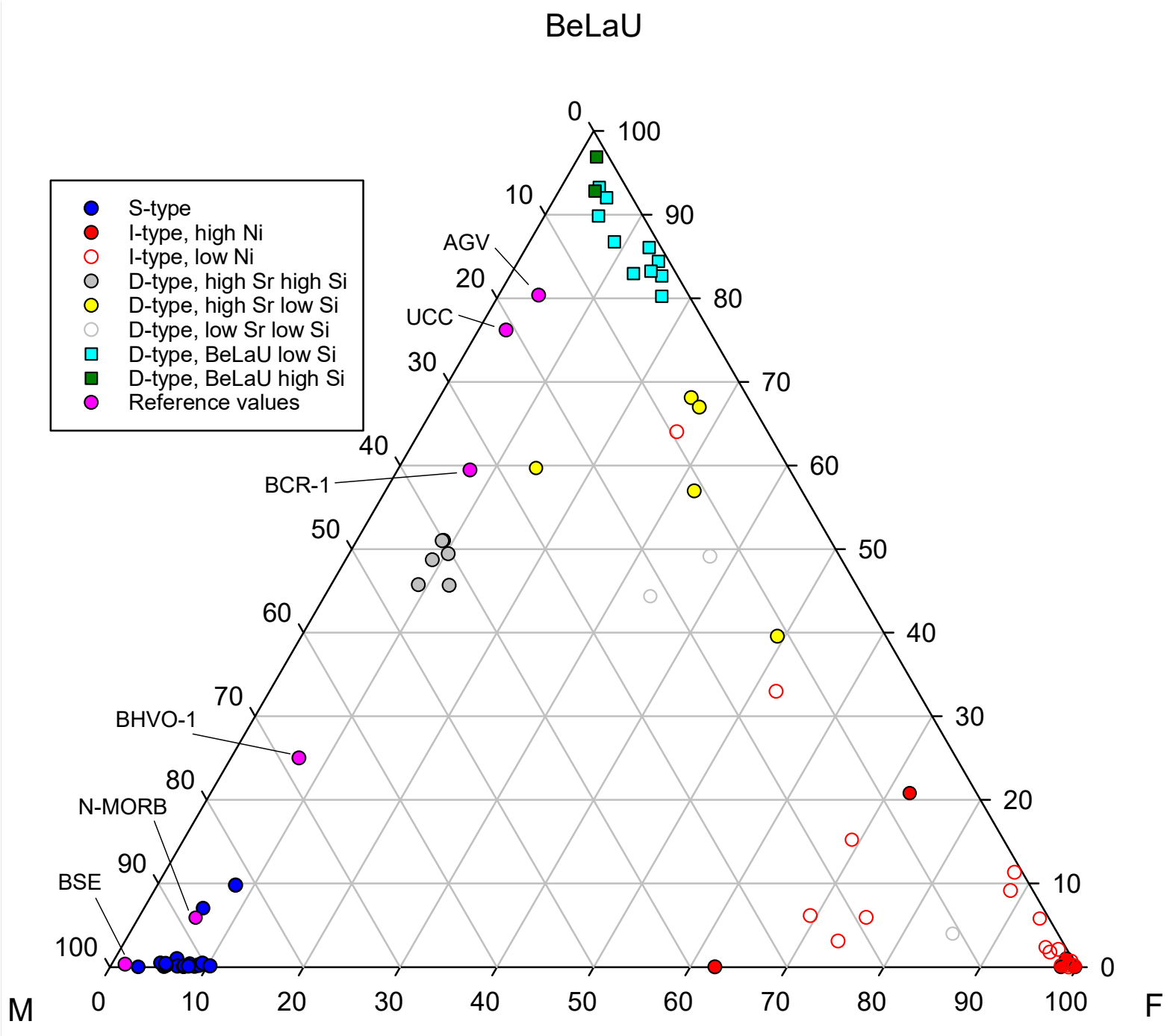

**Figure 13**. Ternary diagram for identifying the spherules that are most enriched in Be, La and U (BeLaU) relative to Mg (M) and Fe (F) in terms of bulk composition. The parameters are CI chondrite normalized (see text).

## Discussion

In total we analyzed 850 magnetic particles by either micro-XRF (792) or ICP-MS (68) or both (11). We found that 48 have simple chemical compositions that likely make them derived from industrial products or terrestrial minerals. They were thus not included in this study and so we are left with 802 spherules for which we report data in this paper. Micro-XRF and ICP-MS data represent bulk compositions, while EPMA data represent spot analyses.

### *Origin of Primitive spherules*

The primitive spherules are in general characterized by high Cr, Co and Ni and highly variable Mn. Chromium is moderately volatile, similarly to Mn, thus in a Cr/Fe vs Mn/Fe plot (**Figure 14**) there is an overall positive correlation as expected for these ratios. We note that S-type spherules have both Cr/Fe and Mn/Fe similar carbonaceous chondrites (CCs) and non-carbonaceous chondrites (NCs), and there is no major effect from volatile loss of Cr and Mn for this type of spherules. Most of the low Ni I-type spherules have normal chondritic Mn/Fe and slightly lower, so these do not have evidence for substantial volatile loss for Mn. This suggests that their low Cr content is due to some other partitioning process. The primitive nature of S- and I-type spherules are supported by their chondrite-like Cr/Fe and Mn/Fe, distinctly different from the planetary values (Earth, Moon, Vesta, Mars) shown in **Figure 14**. In a plot of Co/Fe vs Mn/Fe (**Figure 15a**) when comparing S-type spherules similar CCs and NCs, there is no major effect from volatile loss. When comparing with planetary bodies we see an overall negative correlation that is likely due to core formation in the planets. Ni-rich I-type spherules have very low Mn/Fe showing substantial volatile loss. Low Ni I-type again have normal chondritic Mn/Fe and slightly lower, so these do not have evidence for substantial volatile loss. The primitive nature of S- and I-type spherules are supported by their chondrite-like Co/Fe and Mn/Fe, distinctly different from planetary values. In a plot of Ni/Fe vs Mn/Fe (**Figure 15b**) we also see an overall negative correlation. Here the S-type spherules have lower Ni/Fe compared to CCs and NCs. Ni is more strongly partitioned into metal compared to Co, so some Ni rich metal must have been lost in producing the S-type Spherules (Genge et al. 1996; 2017). Thus, overall Mn, Cr, Co and Ni relative to Fe in S- and I-type spherules supports their origin from primitive chondritic meteorites but do show evidence of volatile loss as well as metal-silicate separation due the processing into spherules.

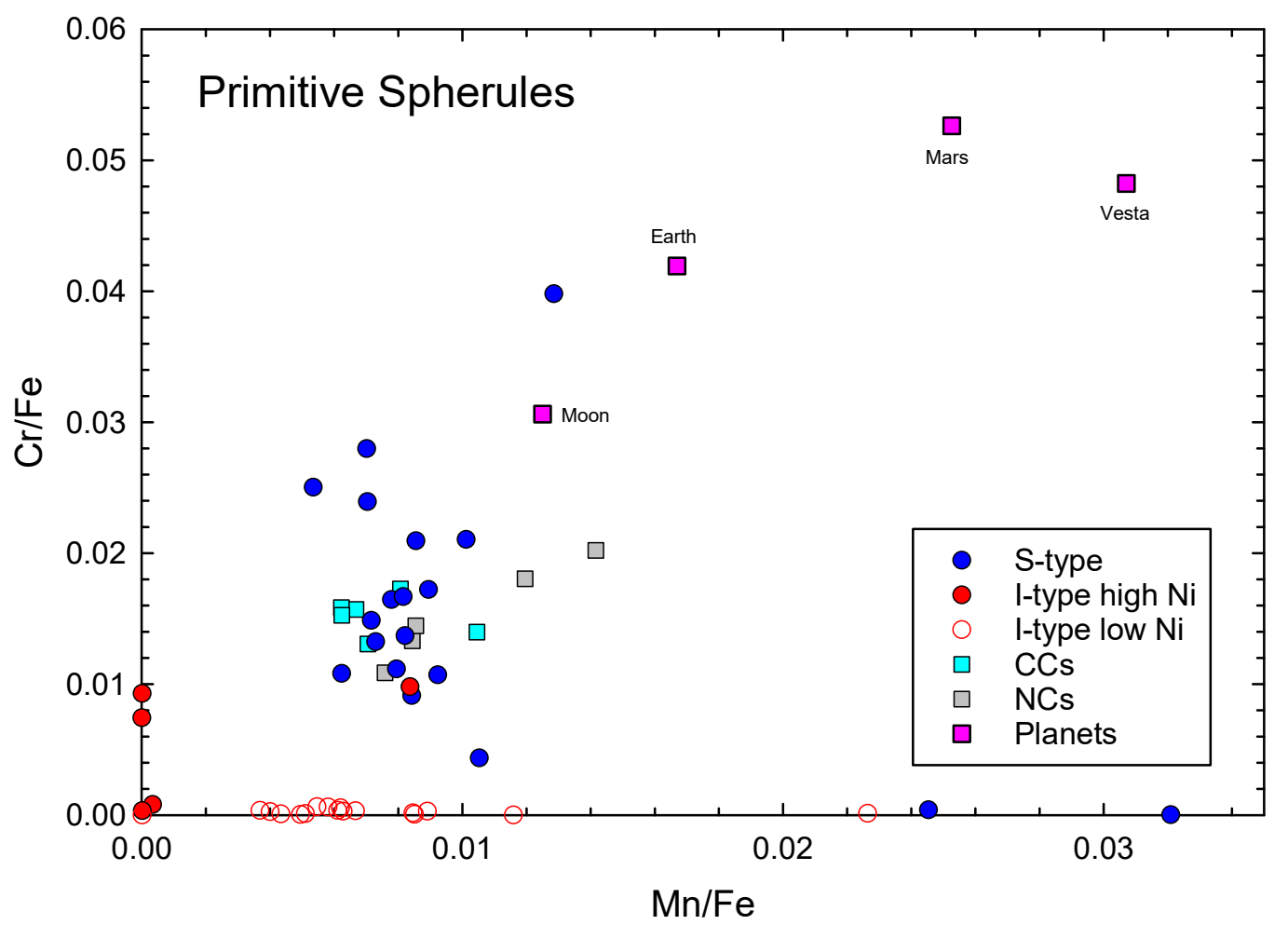

**Figure 14**. A plot of Cr/Fe vs Mn/Fe for primitive spherules compared to carbonaceous (CC) and non-carbonaceous (NC) chondrites, Earth, Mars, Vesta and the Moon. Sources: CCs and NCs from Alexander 2019ab. Mars: Yoshizaki and McDonough (2020), Vesta: Dreibus and Wänke (1980), Moon: Hauri et al. (2015).

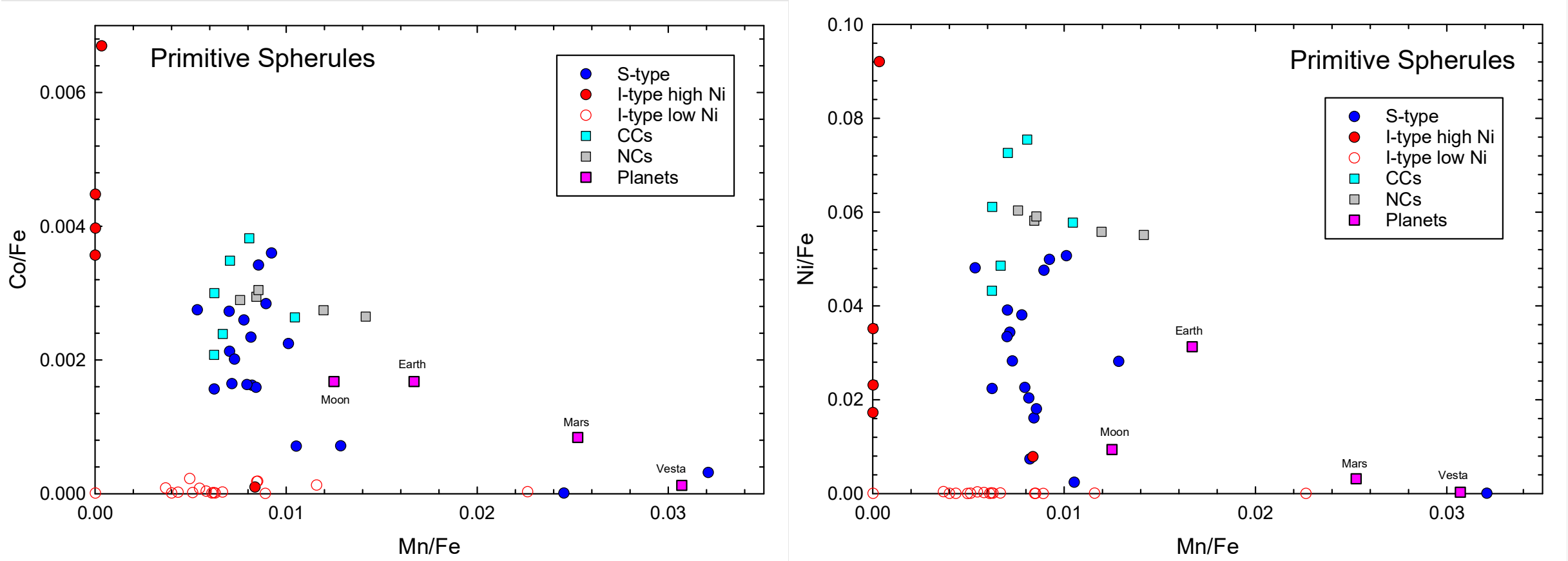

**Figure 15**. A plot of Co/Fe (a) and Ni/Fe (b) vs Mn/Fe for primitive spherules compared to carbonaceous (CC) and non-carbonaceous (NC) chondrites, Earth, Mars, Vesta and the Moon. Sources: CCs and NCs from Alexander 2019ab. Mars: Yoshizaki and McDonough (2020), Vesta: Dreibus and Wänke (1980), Moon: Hauri et al. (2015).

### *Origin of the spherules exhibiting S, I and G-type compositions*

Ablation spheres, which are formed from meteorite fragments that break off from larger meteoroids during atmospheric entry, largely resemble meteorite fusion crusts with respect to their composition and exhibit small differences in comparison to cosmic spherules. Ablation spheres are formed at low altitudes (<40 km) and higher oxygen fugacity conditions than cosmic spherules (>70 km), which undergo evaporative loss under more reducing conditions (Genge et al. 2008). This translates to a difference in enrichments in alkali metal content (such as Na and K) and NiO content in bulk compositions and mineral phases, which has been considered diagnostic in distinguishing the two as higher $f_{O2}$ conditions are favorable in preserving alkali metals during atmospheric entry (Genge and Grady, 1999; Cordier et al. 2011a). While the generally low Ni content, as well as low Na and K contents, suggest the likelihood that the spherules classified as "primitive" in our collection are more akin to cosmic spherules in their formation processes, other micrometeorites in this study that compositionally align with the three classes of cosmic spherules (S-type, I-type and G-type) in the Mg-Si-Fe ternary diagram exhibit relatively high concentrations of alkali metals, in some cases exceeding 2%. This suggests that these materials may be ablation products rather than cosmic spherules that entered the atmosphere at high altitude. However, the dissimilarity to terrestrial materials in a ternary Mg-Fe-Si plot composed of major elements, still indicates that the precursors of these materials were chondritic rather than terrestrial. During the initial stages of EPMA analysis, we observed a NaCl crystal in a pocket of one of the spherules. Some of the high Na contents of a subset of the spherules that were analyzed may include sodium chloride crystals from seawater contamination or may be from alteration.

***Origin of D-type spherules***

D-type spherules may have several origins. The Mg-Al-Fe plot in **Figure 11** has been used for classifying terrestrial komatiites, ultramafic lavas that are primarily found in the Archean (cf. Rickwood 1989). The only modification is that Fe+Ti is replaced with Fe. Such a plot is used in **Figure 16** to further understand the origin of D-type spherules. The importance of this diagram is that it shows komatiites as separate fields from tholeiitic basalts and calc-alkaline and tholeiitic magma series. The blue outline shows the field of common terrestrial rock while the red outline includes all terrestrial igneous rocks. Many of the D-type spherules plot closely to the compositional domains formed by terrestrial igneous rocks. However, all except some of the Si-rich D-type spherules are outside the field of common terrestrial igneous rocks and most of them are outside the red outline where there are no terrestrial igneous rocks, making the interpretations of their origins important.

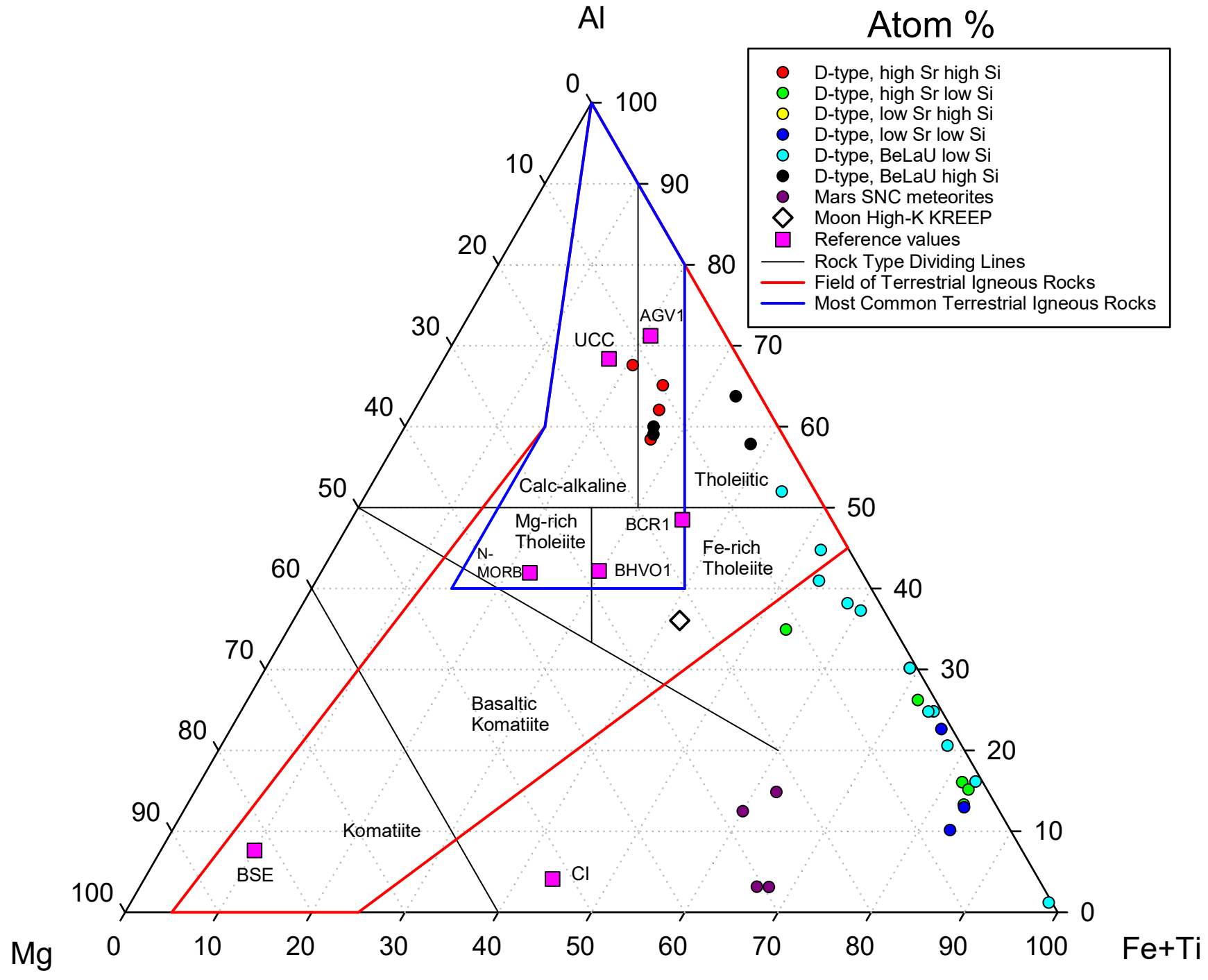

**Figure 16**. Atomic Mg-Al-Fe plot (atom %) Comparison of IM1 site spherules to differentiated materials from Earth (same references as for **Figure 8**), Moon (High-K KREEP; Warren, 1989) and Mars (SNC meteorites; Lodders, 1998).

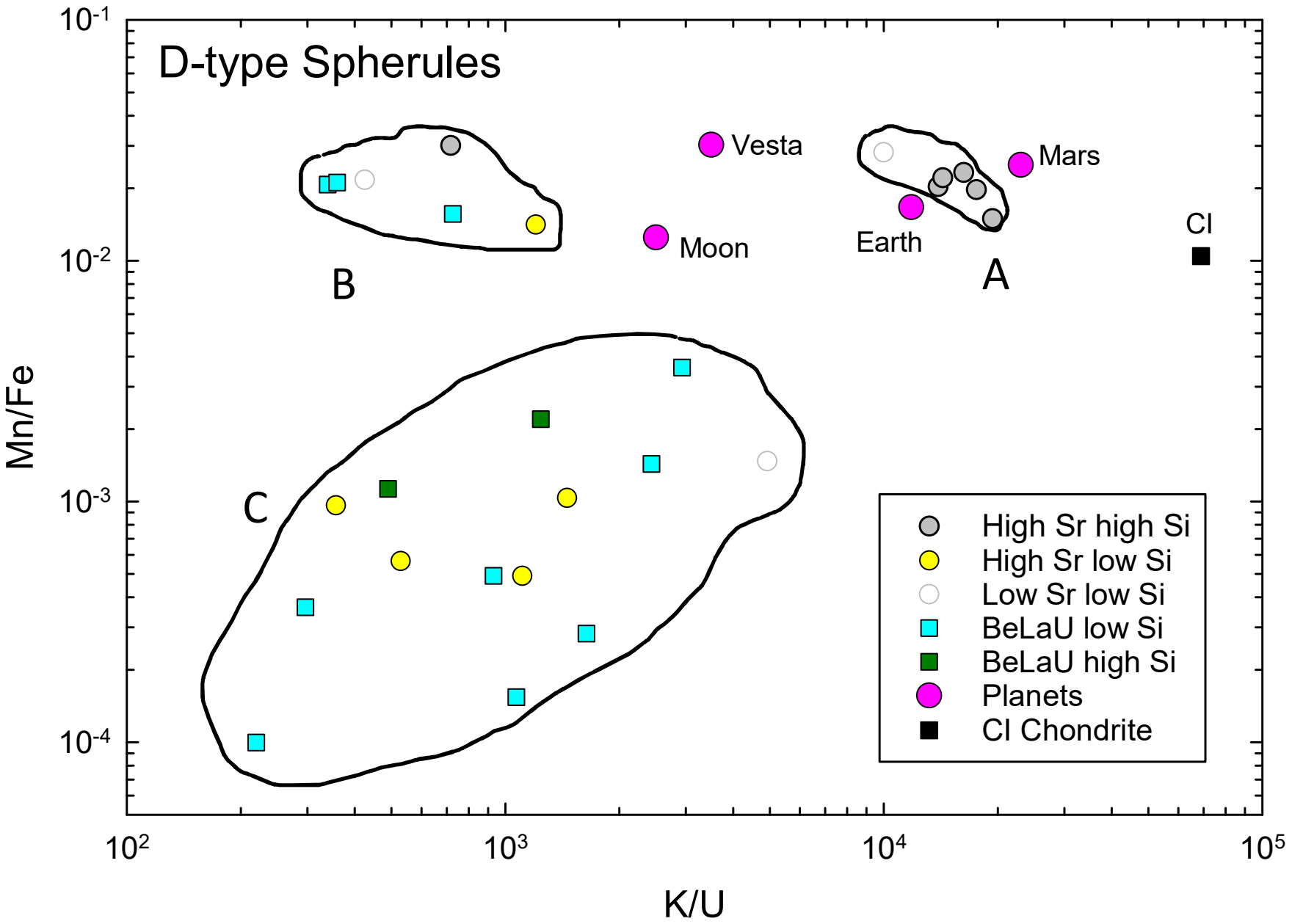

**Figure 17**. A plot of Mn/Fe vs K/U for D-type spherules compared to CI chondrites, Earth, Mars, Vesta and the Moon. Sources: Anders and Grevesse 1989; McDonough and Sun 1995 Mars: Yoshizaki and McDonough (2020), Vesta: Dreibus and Wänke (1980), Moon: Hauri et al. (2015).

***Origin of the novel BeLaU D-type spherules***

The D-type spherules with enrichment of beryllium (Be), lanthanum (La) and uranium (U), labeled "BeLaU," appear to have an exotic composition different from other Solar System bulk rock compositions. The BeLaU spherules are a subset of the D-Type spherules described in the previous section. They are clearly different from the lunar high-K KREEP composition, which is the magmatic composition on the Moon with the highest concentration of Be, La and U. Martian SNC meteorites reflect the Fe-rich composition of the martian mantle and are the only known solar system igneous rocks that plot in the triangle between the red line and the Fe+Ti apex of the diagram. However, there are no martian igneous rock that get as close to the Al-(Fe+Ti) side of the diagram as the BeLaU-type spherules. The BeLaU-type spherules composition could probably only be produced by extreme differentiation of an iron-rich martian-like magma that produced the igneous rocks among the martian SNC meteorite.

BeLaU spherules are observed to be depleted in moderately volatile elements such as Zn, Mn, K and highly volatile elements such as Pb. In addition, K and Mn are both volatile during planet formation while U and Fe are more refractory, so planets end up with different K/U and Mn/Fe ratios. The data for the D-type spherules are shown in a K/U vs Mn/Fe diagram (**Figure 17**). The reason for using these elemental ratios is that they both have for a long time been used to distinguish materials from different bodies in the solar system (e.g., Papike et al. 2017; Halliday and Porcelli, 2001). This works because K and U are both highly enriched in melts, so their ratio does not change much during igneous processes. Therefore, unaltered igneous rocks typically directly reflect a ratio close to the planetary value. In contrast Fe and Mn are both roughly equally distributed between melts and solids, so their ratio also does not change during igneous processes. Since K and Mn are both volatile one would expect both to be lost from the spherules during their flight through the atmosphere. The D-type spherules are compared with primitive chondritic meteorite and planetary values in **Figure 17**. There are three distinct groups. The group A spherules are similar to the Earth values and could by this comparison represent some terrestrial volcanic material. We note that these are all Si-rich spherules which also make them similar in composition to terrestrial igneous rocks. Group B is strongly depleted in K relative to the Earth. Group C is strongly depleted in both K and Mn relative to the Earth. Most of the BeLaU spherules are in the group C field and none are in the Earth-like group A field. This shows that all the BeLaU spherules have exceptionally high volatile element loss.

The shapes of the BeLaU-type spherules in **Figure 4c** and **f** have been seen before in airburst spherules (Tankersley et al. 2024; Van Ginneken et al. 2010; 2021; 2024). A magnetite/spinel rim is a quench feature that is observed in micrometeorites, meteorite ablation debris, airburst particles and impact spherules. This is observed in the two cross-section analyses of two D-type spherules, one of which is determined to be a BeLaU spherule (**Figure 6a, b, c**). Magnetite rims in micrometeorites have been considered to be unequivocal evidence of an extraterrestrial origin (Genge et al., 2008) as these are formed due to hypervelocity deceleration during atmospheric entry (Toppani et al. 2001; Toppani and Libourel, 2003). The dendritic patterns, magnetite rim, as well as the glassy mesostasis, are indicative of quenching.

The results for the "BeLaU" spherule S21 are displayed in **Figures 18a and 18b**. We use this spherule to point out some of the unique feature of the "BeLaU" elemental composition. A plot of the elemental abundances of the spherule S21 (normalized to CI chondrites) as a function of atomic number for 56 elements is shown in **Figure 18a**. Across the diagram the peak abundances are for Be, La and U, hence the name "BeLaU." The abundance pattern of S21 implies derivation from a planetary crust, highly enriched in refractory lithophile elements (red dots). The volatile element abundances (green dots) are very low, suggesting either derivation from an extremely volatile-depleted planet or evaporative loss during passage through the Earth's atmosphere. The very low content of refractory siderophile elements with affinity to iron (Re) suggest a source planet with an Fe core. Since there are strong indications that the spherules are derived from a differentiated planet, the data are also plotted in **Figure 18b** as a function of an igneous compatibility sequence. Compatibility is a geochemical parameter measuring how readily a particular element substitutes for a major element in mantle source minerals during melting to produce magma. It also roughly represents the sequence of enrichments of elements in a crystallizing magma. In general, it is not clear how representative an individual spherule will be of its host meteorite. However, with respect to BeLaU, we find an elemental pattern that is repeated throughout multiple spherules. In addition, the smooth elemental abundance patterns when arranged with respect to their compatibility, and lack of features indicating a strong presence of individual minerals such as Be- and Th-silicates, uraninites, and K-feldspar suggest that whatever process that produced the spherules by vaporization and melting produced materials that resemble igneous rocks, rather than igneous minerals. Our chemical analysis suggests a magmatic origin of such samples as opposed to minerals or nugget effects from minerals that would dominate elemental abundance patterns. These elemental abundance patterns likely preclude anthropogenic origins for BeLaU.

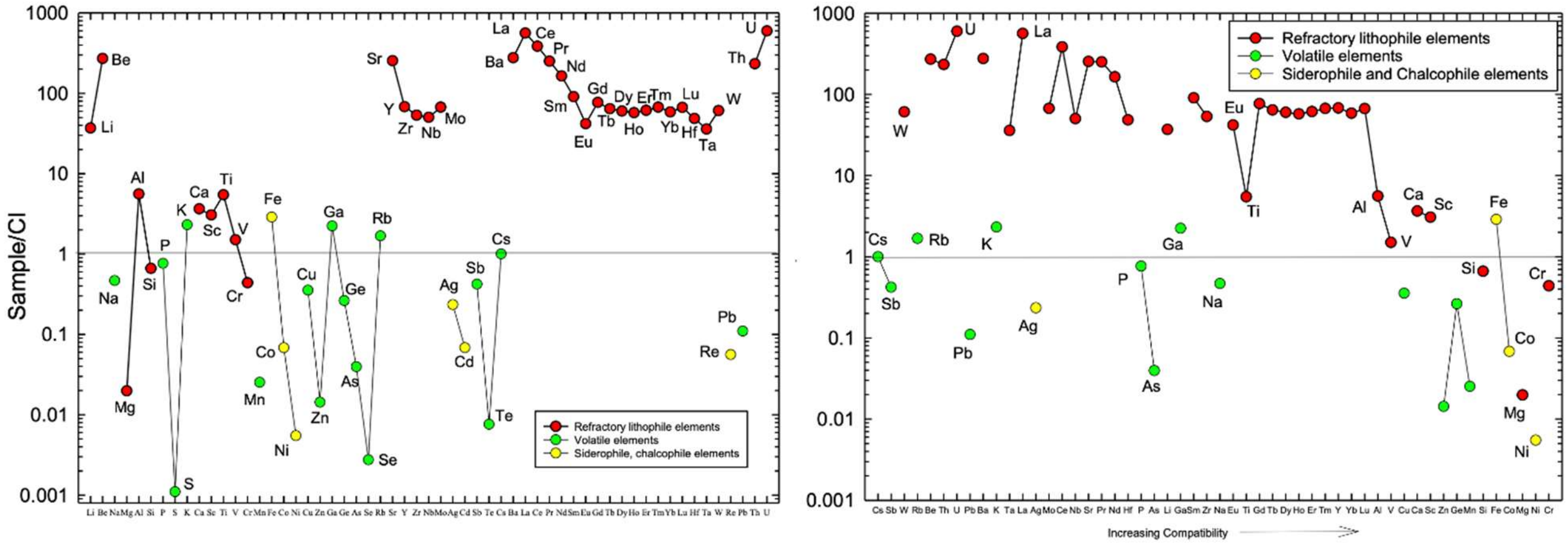

**Figure 18**: a) The "BeLaU" elemental abundances of the spherule S21 (normalized to CI chondrites) versus atomic number for 56 elements. The solar system standard of CI chondrites is represented by a value of unity on the plot. b) Abundances versus compatibility (see text) for S21. Elements ordered with increasing compatibility towards the right.

## Further Discussion on D-Type Spherules and the BeLaU Subset

### *Exploring a terrestrial origin for the D-type spherules*

Five of the D-type spherules plot in the domain delineated by the most common terrestrial igneous rocks in a Mg-Al-Fe ternary diagram (**Figure 11b**, grey circles). These five spherules also plot close to the rocky planetary bodies (the Moon, Mars, Earth and Vesta) in terms of their K/U and Mn/Fe ratios which suggest that these spherules may be terrestrial in origin (**Figure 17**). Therefore, it is suggested that terrestrial processes may have well been involved in forming a subset of the D-type spherules in this collection. About 20% of the spherules classified as D-types in this study to a large degree corresponding to the "high Si, high Sr" subclassification, overlap in composition with respect to the most common terrestrial igneous rocks in the Mg-Fe-Si ternary diagram, also suggesting that these spherules may have terrestrial origins. However, the Mn/Fe and K/U ratios indicate that a substantial fraction of these D-type spherules analyzed through ICP-MS is dissimilar from terrestrial values (**Figure 17)**. We provisionally refer to these as "nonterrestrial" to acknowledge that there is no known terrestrial analog to attribute their origins.

Volcanism may be one such possible mechanism for generating spherules exhibiting a highly differentiated element abundance. Yet volcanic spherules are understudied, and the full extent of the mechanisms involved in generating such spherules are unclear. Their origins so far are largely classified to be from i) low viscosity magmas (Porritt et al. 2012) or ii) hydrovolcanism (Amonkar et al. 2021; Agarwal and Palayil 2022). Spherules that are suggested to have a hydrovolcanic origin are noted to be larger in size than spherules formed from volcanic ash, ranging up to millimeters in diameter. In discerning spherules of volcanic origins from extraterrestrial origins, elemental compositions have been used. In particular, Mn has been used as an indicator of hydrovolcanic origin (Finkelman, 1970; Agarwal et al. 2022) owing to the inferred low oxygen fugacity conditions needed to host Mn in the magnetite phase and the lack of observed abundances in cosmic spherules. While the D-type spherules may still be terrestrial in origin, the low Mn content of many of the D-type spherules analyzed through ICP-MS in comparison to known planetary bodies (**Figure 16**) indicate a strong depletion and thus likely eliminate hydrovolcanism as an origin, although volcanic processes are difficult to fully preclude at this time.

The enrichment of elements such as Ti and Al have also been used as an indicator for discerning spherules with a terrestrial as opposed to an extraterrestrial origin (Chang et al. 2017; Iyer et al. 1997; Finkelman, 1970; Fredrickson and Martin, 1963). While most if not all cosmic spherules have been considered chondritic or achondritic in origin, in contrast to meteorite collections, cosmic spherules from more differentiated planetary bodies such as the Moon or Mars have yet to be widely discerned. Planetary bodies that have undergone differentiation exhibit signs of fractionation and are enriched in incompatible elements. Thus, it is not clear that enrichment of elements such as Ti or Al, or REE patterns can be used as criteria for distinguishing terrestrial precursors from micrometeorites of extraterrestrial origin. For example, such signatures would be expected for spherules derived from highly differentiated lunar or martian meteorites.

Further comparison of the bulk composition of D-type spherules with terrestrial magma compositions are shown in **Figure 19** for the two most abundant elements (Fe and Si as oxides) in the D-type spherules. Most terrestrial magma compositions roughly follow the red line between basalt and granite in this diagram. Relatively few D-type spherules are close to this trend. Most of them fall in a triangle between magnetite and the terrestrial magma trend as they have abundant magnetite. Silicate mantles for other solar system or extra-solar system mantles could be more iron rich but not likely to produce basalts with more than 20–25 % FeO as found in some basalts from Mars, Moon and Vesta (Warren, 1989; Lodders 1998; Jambon et al. 2002; Kitts and Lodders 1998). Fractional crystallization of such basalts would also trend towards the granite composition. The D-type spherules were obtained by our sampling method because they are rich in magnetite as shown in earlier EPMA images of some of the D-type spherules (**Figure 6**), some present as relict grains. The triangular area with the data in **Figure 19** is consistent with this explanation. Thus, the bulk compositions of the spherules do not, in most cases, represent magma compositions but are likely mixtures of silicate magma compositions and magnetite. Magnetite has very low concentrations of most other elements (e.g., Dare et al. 2014) and will therefore primarily act to dilute most elements equally, but not substantially alter the trace element pattern of the melt. In general, the BeLaU incompatible element concentration pattern discussed in the following will represent minimum values. The true BeLaU trace element composition of the melt will be even higher than the values shown in the diagrams.

In **Figure 20,** we show a $FeO_{1.5}$–$SiO_2$ plot for EPMA data of polished sections of D-type spherules, showing spot analyses for the most common minerals as well as the interstitial glass. The most abundant mineral grains are magnetite ($Fe_3O_4$) with a small to in some cases substantial hercynite ($FeAl_2O_4$) component. A few grains of hercynite were also identified. This plot confirms the main conclusion from the bulk spherule data in **Figure 19** as the interstitial glass is close or somewhat above the terrestrial silicate melt trend and the mineral grains concentrated close to the magnetite composition. Thus, the intermediate compositions seen in **Figure 19** and largely absent in **Figure 20** must be mixtures of a magnetite solid solution with a melt composition that is similar to but not necessarily the same as for terrestrial igneous rocks. In fact, many of the glass compositions in **Figure 20** would be easier to explain as coming from a mantle that is more iron-rich than the Earth's mantle.

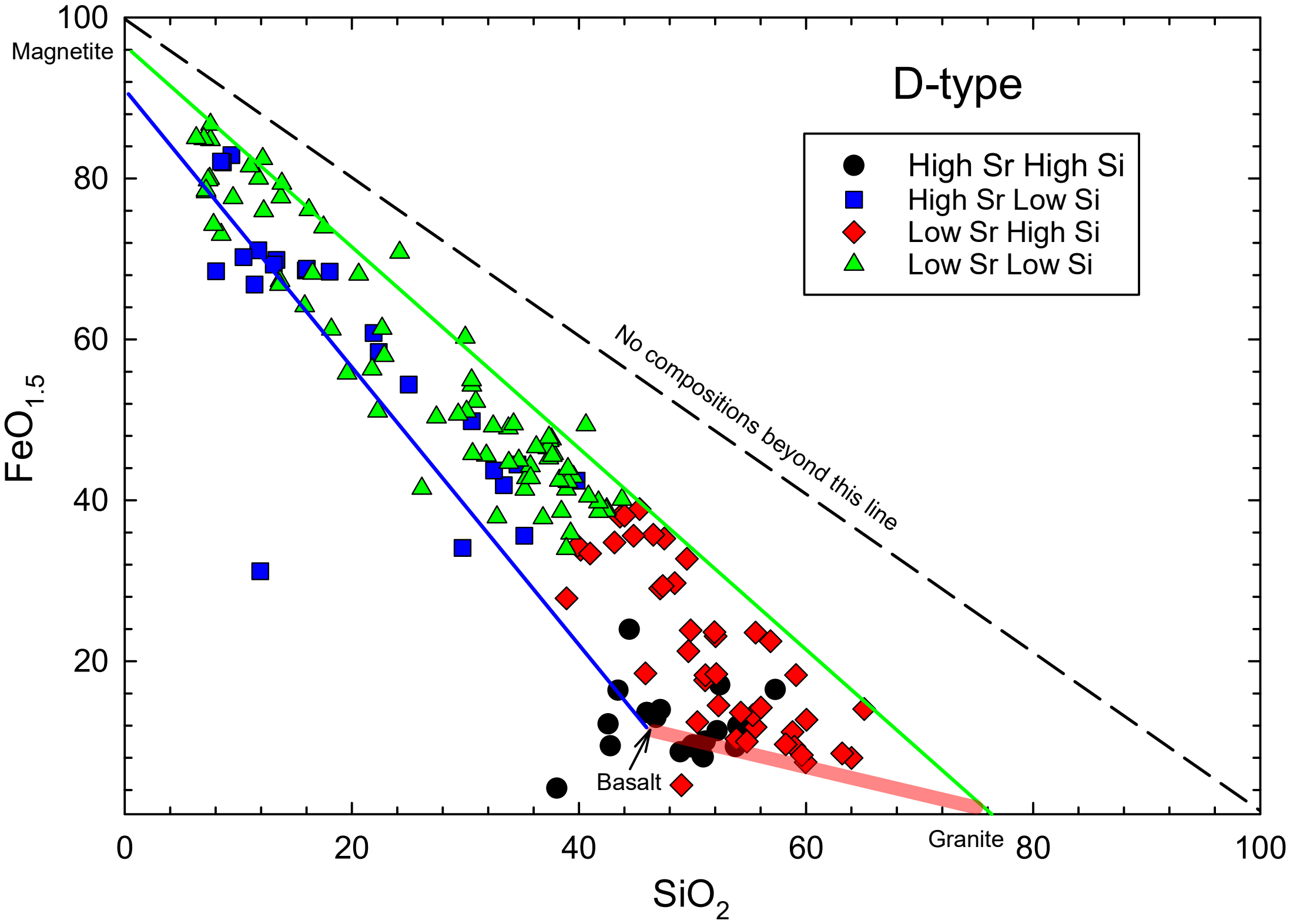

**Figure 19.** $FeO_{1.5}$–$SiO_2$ plot (bulk composition, weight %) of micro-XRF data for the various classes D-type spherules. Comparison with the typical trend in the composition of terrestrial igneous rocks (e.g., Winter 2001), shown as a red line between basalt and granite. The position of magnetite is also shown and that most of the data lies within a triangle enclosed by the compositions of basalt, granite and magnetite.

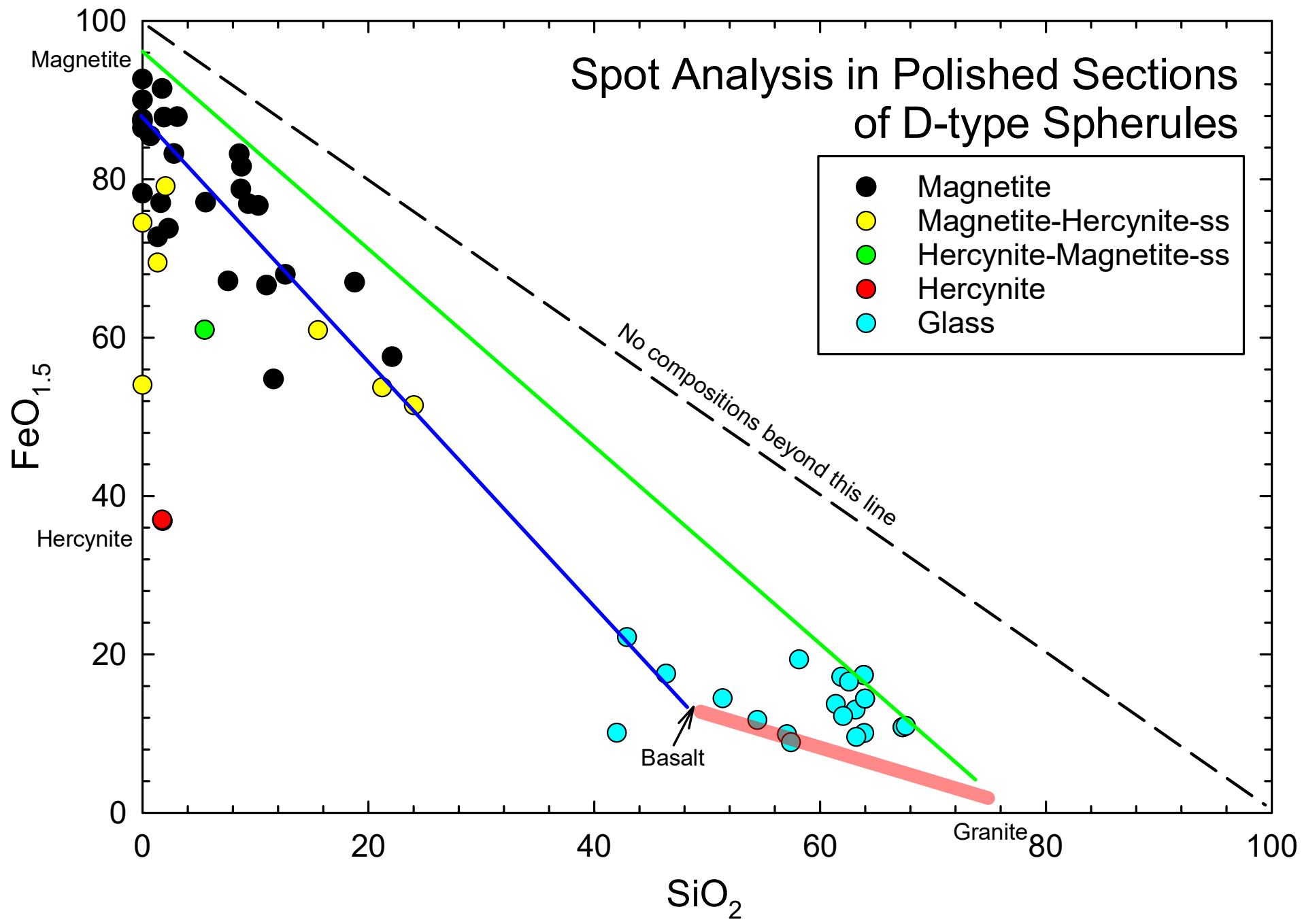

**Figure 20.** $FeO_{1.5}$–$SiO_2$ plot (weight %) of EPMA data of polished sections of the two D-type spherules analyzed in this study, from spot analyses for the most common minerals as well as the interstitial glass. Data are in **Table S4** and images of all the spot analyses are in **Figures S4 S5, S6, S7, S8, S9, S10, S11 and S12**.

***Comparisons with Australasian microtektites***

As spherules are small fragments of their precursors, it may be difficult to infer the bulk rock composition of the parent body from analysis of these alone. However, abundance patterns may serve as identifying traits of the parent body or melted precursor. Among impact produced spherules of terrestrial origin are tektites and microtektites. Impact spherules are suggested to have formed as melt-ejecta droplets of the upper continental crust (Folco et al., 2009; Glass et al., 2004; Brase et al., 2021), rarely containing contamination from the impactor (Glass and Simonson, 2010).

Bulk compositions of spherules have been previously used to compare with existing groups or trends of microtektites and tektites (e.g., Brase et al 2021, Folco et al. 2009). Given that microtektites and tektites from a strewn field are characterized to be homogeneous (Koeberl, 1994), such bulk analysis may still be somewhat diagnostic of their origins and instrumental in supporting or eliminating such sources as their origins. Normalized atomic Mg-Fe-Si bulk compositions of Australasian microtektites delineate a distinct domain in a ternary plot (**Figure S3**) which overlap with most common terrestrial igneous rocks. The strewn field associated with Australasian microtektites covers a wide geographic area, and its known borders extend from Madagascar, Antarctica, parts of West Papua, Indonesia, to as far as eastern Australia (Folco et al., 2023). Thus, any spherules in our collection which are identified to be microtektites are likely to be closest to the Australasian microtektites in composition. Comparisons of spherules and Australasian microtektites plotted with respect to normalized atomic proportions of Mg, Fe and Si indicate partial overlap with a subset of D-type spherules, suggesting that these spherules may indeed be related to the Australasian microtektites. However, the BeLaU spherules demonstrate a distinct enrichment in Fe and do not overlap with this domain in the Mg-Fe-Si ternary plot in terms of composition.

Many microtektites display distinct characteristics that distinguish them from cosmic spherules in certain aspects of their morphology and mineralogy, which include concentric zonations and radial patterns in polished cross sections, lechaterlierites, or smooth, homogeneous glassy textures, particularly if they have been ejected more than 10 crater diameters away from the source (Glass and Simonson, 2012). Although a fraction of the spherules in our study seem to indicate a similarity in terms of composition to the aforementioned microtektites, we exercise caution in solely attributing these to the Australasian microtektites until further characterizations have been made. Further detailed studies of these spherules are anticipated to help present a comprehensive picture on the origin of the D-type spherules in this study.

***Comparison of BeLaU pattern with terrestrial igneous rocks***

The abundances of elements as function of their compatibility (continental crust compatibility sequence– see below) for all 12 "BeLaU"-type spherules are shown in **Figure 21**. The "BeLaU" spherules' variations in the abundances of trace elements relative to CI chondrites are higher by 1–3 orders of magnitude compared to cosmic spherules from the solar system reviewed by Folco and Cordier (2015). BeLaU samples have prominent negative anomalies of Ti, Li, higher Lu/Al, and variable Be and Sr enrichments that do not match the smooth pattern of the upper continental crust. These are probably a result of crystal fractionation, as for example Ti is depleted once ilmenite or Ti-magnetite fractionation occurs.

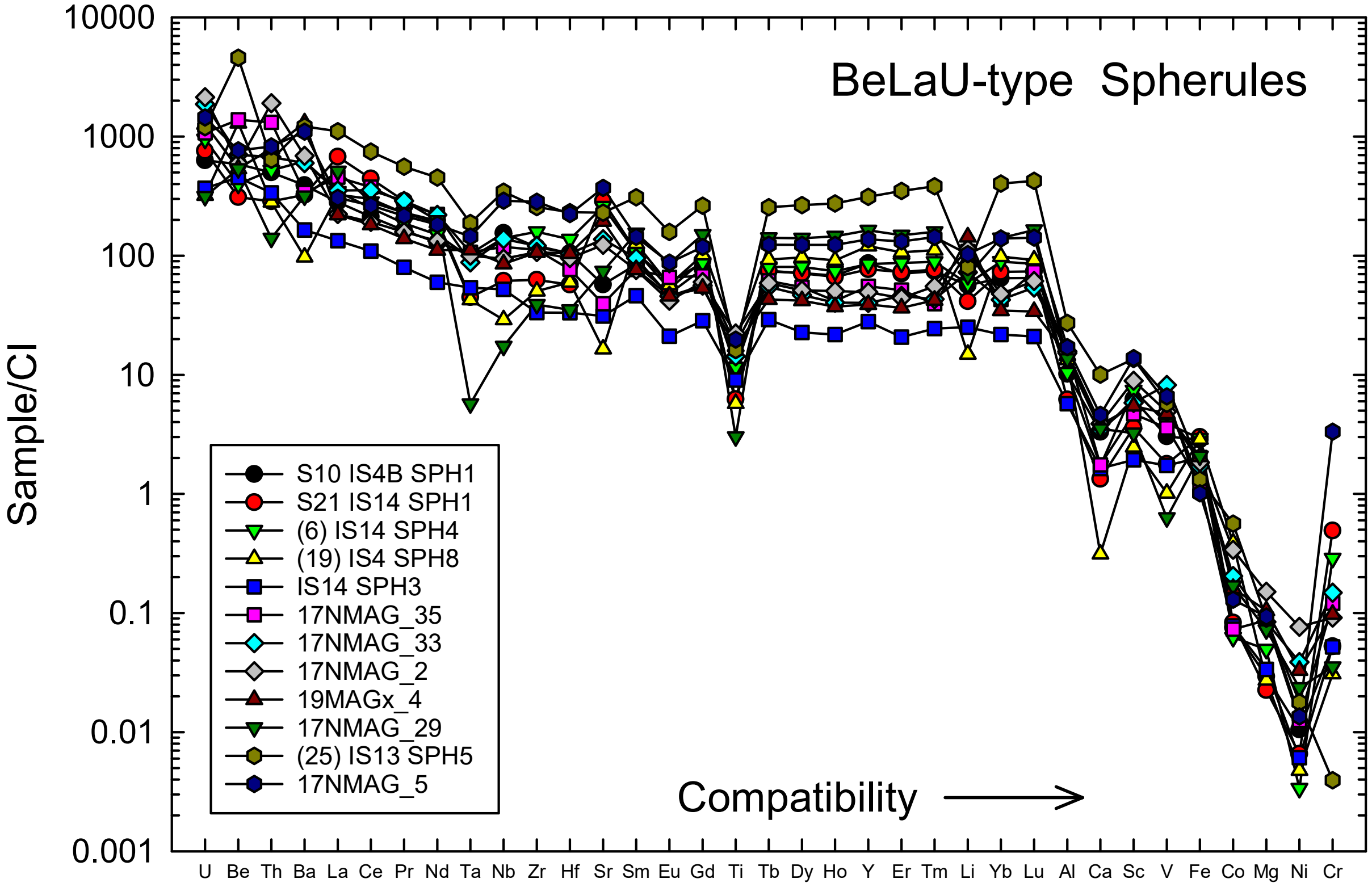

**Figure 21**. Abundances of refractory lithophile elements as function of their compatibility (continental crust compatibility sequence) for all 12 "BeLaU"-type spherules.

The rare earth element pattern of the BeLaU-type spherules is compared in **Figure 22a** with terrestrial igneous rocks. Their concentrations are normalized to the bulk silicate Earth (BSE) and the elements are ordered with increasing atomic number. The parenthesis around Pm is to indicate that this is an interpolated value since Pm does not occur in nature. This element is included in the plot to avoid the appearance of a break in the slope of the REE pattern due to a missing atomic number if Pm is left out of the plot.

The average of the BeLaU elemental pattern is compared with mantle-derived igneous rocks representing the entire terrestrial range from the least enriched being mid-ocean ridge basalts (Hofmann 1988; Sun and McDonough, 1989), through the more enriched ocean island basalts (Hofmann 1988; Sun and McDonough, 1989) to kimberlites which are the most enriched mantle-derived igneous rocks (Dawson and Hawthorne, 1973; Dawson, 1980; Arndt, 2003; Giuliani et al., 2020; Tappe et al., 2020). The average elemental abundance for BeLaU is used to avoid dependence on one spherule to derive broader conclusions (**Figure 21 o**ffers a sense for the variation between the 12 samples used to compute this average). These comparisons demonstrate that none of these mantle-derived igneous rocks are a good match for the BeLaU pattern. In the case of mid-ocean ridge basalts (MORB) this is because of their relatively flat but slightly light rare earth element (LREE – La through Sm) depletion pattern. Ocean island basalts (OIB) are high in LREE but not as high as BeLaU but are much lower than BeLaU in heavy rare earth elements (HREE – Gd through Lu). The most LREE enriched magmas are kimberlites with LREE enrichments similar to those of BeLaU but with much lower HREE content. Their low HREE content ocean island basalts and kimberlites are well understood to be due to garnet in their mantle source retaining the HREE in the mantle (Kay and Gast, 1973). Thus, there is no known way to obtain the BeLaU pattern as a mantle-derived melt on the Earth.

The REE pattern of the Earth's upper continental crust (UCC) (Taylor and McLennan, 1985; Rudnick and Gao, 2014) is very well established. It is the most common REE pattern found both in andesitic rocks of volcanic arcs giving rise to the andesite model for the composition of the continental crust (Taylor and McLennan, 1985). The origin of such magmas involves both ambient mantle sources varying from more depleted than MORB to OIB-like with a relatively constant proportion of subducted oceanic crust material (hydrated basalts and sediment) added to the mantle to produce volcanic arc magmas (Turner and Langmuir, 2022). We note that the shape of the BeLaU pattern is similar to the volcanic arc or UCC pattern, except that the BeLaU pattern is enriched by about a factor of 4 in LREE and a factor of 10 in HREE. While such enrichment could in principle be produced by crystal fractionation, the most enriched rhyolitic magmas in volcanic arcs do typically not display such extreme enrichments (e.g,. Heywood et al. 2020).

It has long been known that rare rhyolites from Iceland have REE patterns similar to that of the UCC but much higher concentrations (O'Nions and Gronvold, 1973), some of the most enriched being from Torfajokull. The rhyolites plotted in **Figure 22a** are from Torfajokull in Iceland (Martin and Sigmarsson, 2007). Such rhyolites are volumetrically insignificant. Iceland and UCC have very similar REE patterns except for the absolute enrichments. BeLaU and Iceland rhyolites are very similar, essentially the same in LREE enrichments and a factor of 2 difference in HREE. Considering the variability in both BeLaU spherules and rhyolites, this is not a significant difference. The Torfajokull rhyolites are best explained by partial melting of hydrated metabasaltic crust producing a magma with REE composition similar to the UCC followed by extreme fractional crystallization of feldspar (Martin and Sigmarsson, 2007). On the basis of the REE patterns, this type of origin for the BeLaU pattern is a possibility that will now be tested with additional elements.

A sequence of increasing compatibility of elements in the source was defined in two independent ways by Hofmann (1988). The first was the ordering of the elements of decreasing BSE-normalized concentrations in the continental crust. The second was based on the BSE-normalized concentration correlations in oceanic basalts, which is more appropriate for mantle melting. However, the two resulting compatibility sequences are surprisingly similar except for Nb, Ta, and Pb. **Figure 22b** shows 41 elements ordered in terms of their increasing compatibility sequence in the mantle. Overall, the relative positions of these elemental patterns are similar to what is seen in the REE diagram. The major difference is between BeLaU and the Icelandic rhyolites. The elements Ba, P, Sr, Ti, Ca and Mg are all much lower in the Iceland rhyolites compared to the BeLaU pattern. This is due to the extreme fractional crystallization of feldspar and some accessory phases (apatite, ilmenite) that are needed to produce the rhyolite composition of the Torfajokull rhyolites. Thus, in addition to such rhyolites being extremely rare on the Earth, they do not match the BeLaU pattern when more elements than the REEs are being considered. We conclude that we do not know of any Earth-derived material with the BeLaU pattern.

The BeLaU BSE normalized, bulk elemental pattern is compared with the BSE normalized average upper continental crust composition in **Figure 23**. This shows that BeLaU-type spherules are depleted by a factor of 2 to more than 10 for all volatile elements (Na, K, Mn, Zn, Rb, Cs, Tl, Pb and Bi), while refractory lithophile elements are enriched by a factor of 3-20 compared to the average continental crust. Also, the rare earth element pattern is substantially fractionated when compared to the average upper continental crust. We must conclude that the elemental pattern of the BeLaU spherules is very different from the upper continental crust composition. The volatile elements (such as K, Mn, Zn and Pb) were possibly lost by evaporation during the passage of a bolide like IM1 through the Earth's atmosphere.

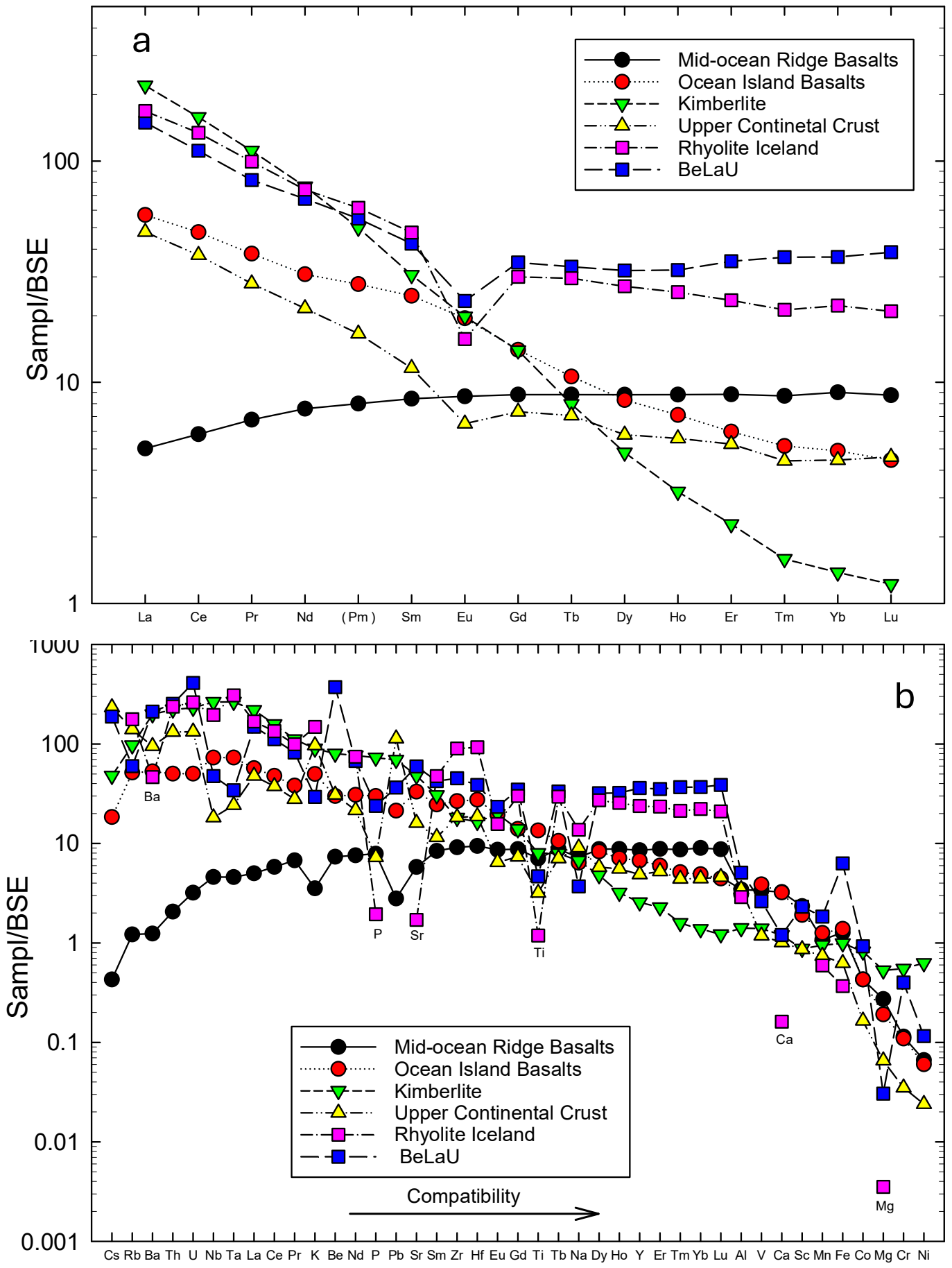

**Figure 22**. a) Comparison of the average "BeLaU" rare earth element pattern (12 samples) with terrestrial volcanic rocks and the average upper continental crust. Elemental abundances are normalized with respect to the BSE for direct comparisons with terrestrial materials. Elements are ordered with increasing atomic numbers. Pm does not occur in nature but is interpolated between Nd and Sm to avoid a break in the slope in the REE pattern due to the lack of this element. b) Comparison of the "BeLaU" compatibility element pattern with terrestrial volcanic rocks and the average upper continental crust. Elements are ordered with increasing compatibility in the mantle source (see text). Data sources: Earth's upper continental crust (Rudnick and Gao, 2014); kimberlites (Dawson 1980; Arndt, 2003; Giuliani et al., 2020; Tappe et al. 2020); mid-ocean ridge basalts and ocean island basalts (Hofmann, 1988; Sun and McDonough, 1989). Average rhyolite from Torfajokull, Iceland (Martin and Sigmarsson, 2007).

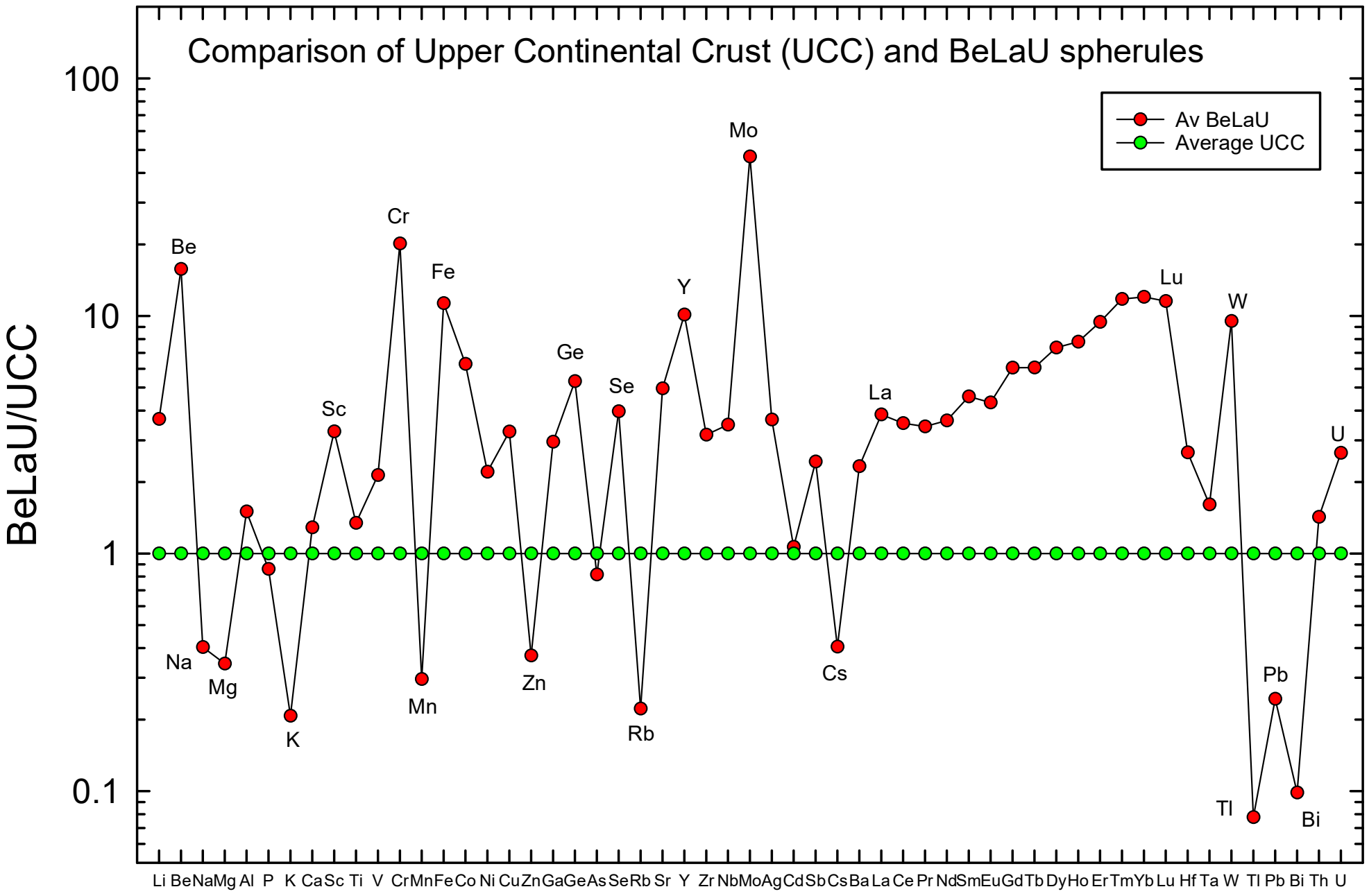

**Figure 23.** Comparison of average BeLaU-type spherules with average upper continental crust**.**

## Exploring Possible Extraterrestrial Origin for BeLaU-type Spherules

*Comparison of BeLaU pattern with extraterrestrial igneous rocks*

Spherules identified as BeLaU-type are compared in **Figure 24** with selected incompatible element enriched extraterrestrial igneous rocks from Mars, Vesta and the Moon. The enriched shergottites (Shergotty, Zagami, and Los Angeles) from Mars (Lodders, 1998; Jambon et al., 2002) have much lower concentration than the BeLaU pattern. The same is true for the eucrites (Bouvante, Pomozdino, and Stannern) from Vesta (Kitts and Lodders, 1998). Thus, there is no indication of an affiliation of the BeLaU pattern with igneous processes on Mars or Vesta.

The most incompatible element enriched composition on the Moon is the well-known KREEP composition (Warren 1989). The KREEP name is an acronym for the incompatible elements K, rare-earth elements (REE), and P. The Moon's anorthositic crust is widely attributed to flotation of plagioclase in a Moon-wide magma ocean. The fractional crystallization of this magma ocean would lead to large enrichments of incompatible elements in a residual melt at the end stage of lunar magma ocean crystallization (Warren and Wasson, 1989). The lunar KREEP composition is widely believed to be the result of crystallization of an early lunar magma ocean (Snyder et al., 1992; Rapp and Draper, 2018; Charlier et al., 2018). This is considered to be the most plausible source of KREEP. As shown in **Figure 24a** and **24b,** KREEP is the most similar extraterrestrial material to BeLaU in composition. However, there are also some distinct differences between KREEP and BeLaU. The much smaller Eu anomaly in the REE pattern of BeLaU compared to KREEP show that the oxygen fugacity of the BeLaU source mantle was similar to Earth's mantle while the Moons mantle had much lower oxygen fugacity close to the iron-wustite buffer (Weill and Drake, 1973; Drake 1975). The similarity to KREEP for many other elements may hint at a magma ocean origin for BeLaU, but it is not likely to be the lunar magma ocean.

These findings suggest that the BeLaU pattern represents an unusual cosmo-chemical signature not explained by currently understood Earth or solar system processes, warranting further investigation into its origin.

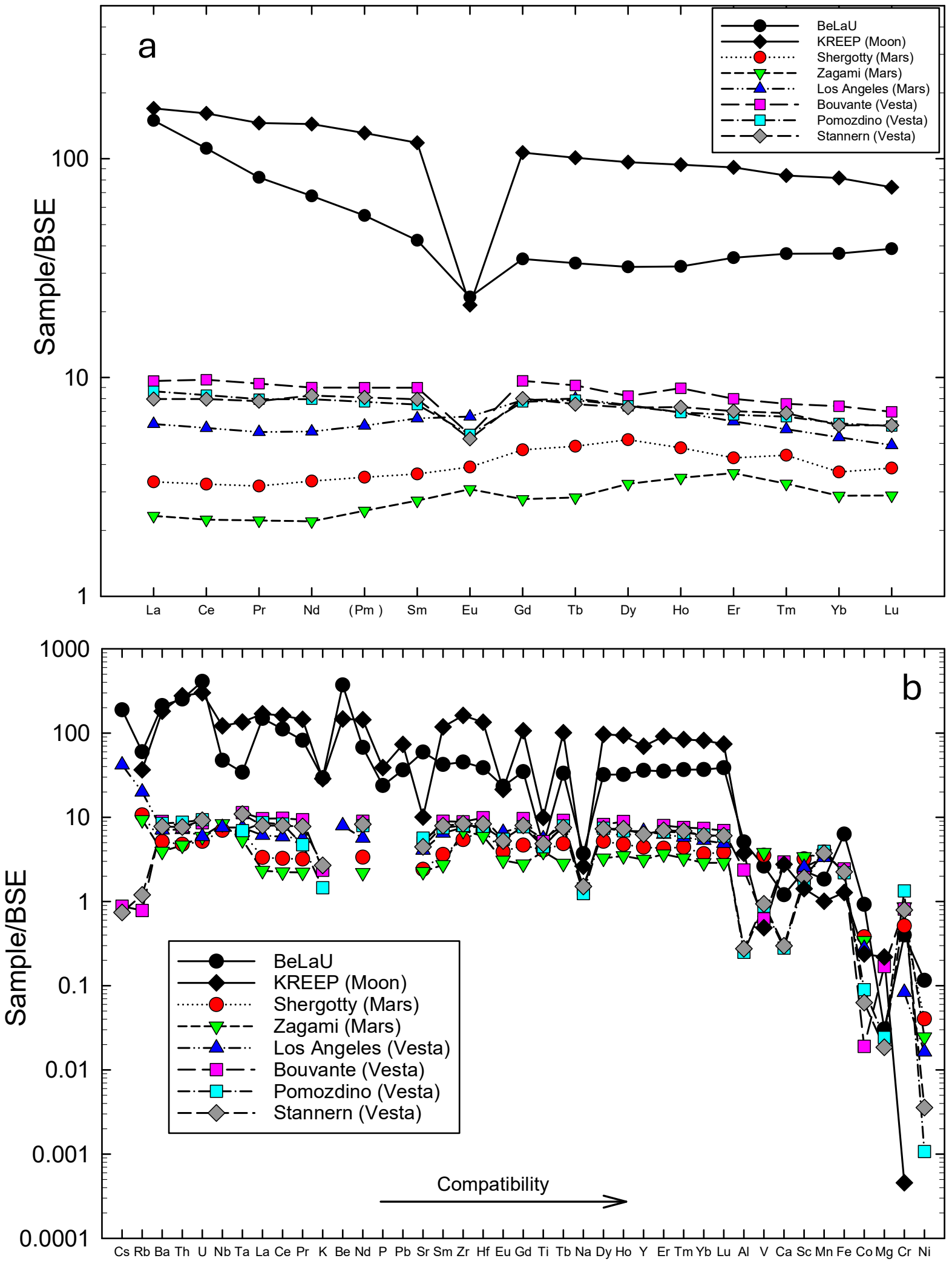

**Figure 24**. a) Comparison of the average "BeLaU" element pattern with terrestrial volcanic rocks and the average upper continental crust. b) Comparison of the "BeLaU" element pattern with extraterrestrial solar system igneous rocks. Elements are ordered in increasing compatibility. For data sources: lunar KREEP (Warren, 1989); shergottites (Shergotty, Zagami, and Los Angeles) (Lodders 1998; Jambon et al., 2002); eucrites (Bouvante, Pomozdino, and Stannern) (Kitts and Lodders, 1998).

*Comparison of BeLaU spherules to anthropogenic materials* It has been claimed that the compositions of BeLaU spherules are consistent with coal ash (Gallardo, 2023). The National Institute of Standards and Technology (NIST) has provided standards of coal fly ash. The best documented standard for many elements is SRM 1633a (Jochum et al. 2005). We compare the average composition of BeLaU spherules for 55 elements with the SRM1633a coal fly ash standard in **Figure 25**. Many volatile elements (Zn, As, Se, Cd, Tl, Pb and Bi) are enriched in the coal fly ash by factors of about 10 to 100 compared to the BeLaU spherules. Some refractory elements (Be, Ca, Cr, Fe, Y, Tm, Yb, Lu W) are depleted by factors of 3 to 10 in coal fly ash when compared to BeLaU spherules. Thus, BeLaU spherules do not have the composition of coal fly ash, making the claim of Gallardo (2023) invalid (Loeb et al. 2024a). The BeLaU composition shows an excess of Be, La and U, and other elements by up to three orders of magnitude relative to the solar system standard of CI chondrites.

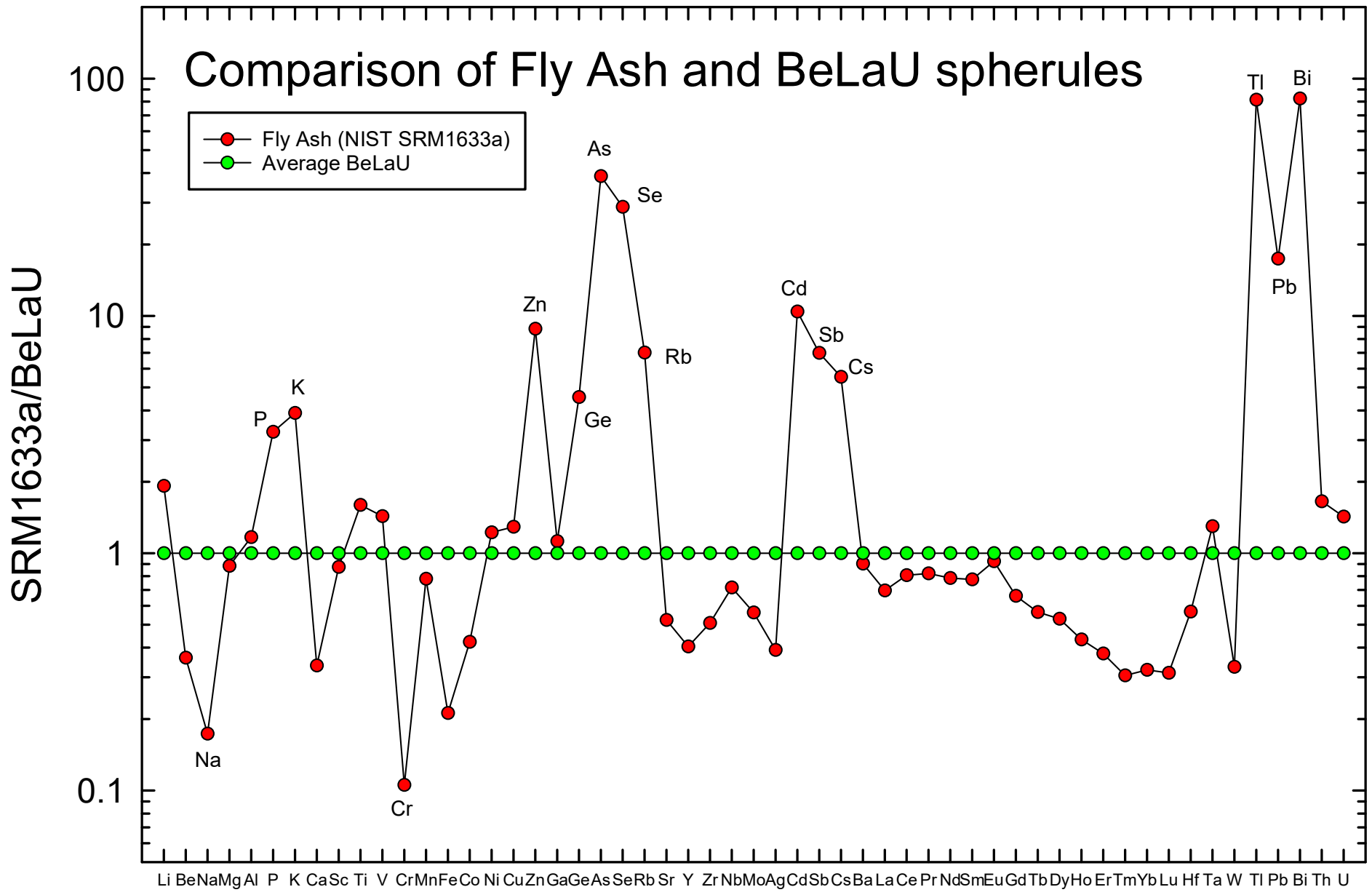

**Figure 25.** Comparison of BeLaU with the NIST coal fly ash standard SRM1633a for 55 elements**.**

**Conclusions**

The magnetic sled survey around IM1's fireball site during 14–28 June 2023 resulted in 850 magnetic particles (consisting mostly of spherules) of diameters within the range of 0.1–1.3 millimeters through 26 runs surveying 0.06 $km^2$. It is not yet clear as to whether any of these specimens are associated with the IM1 bolide. From elemental analyses, we identified four distinct groups of primitive spherules (S-type, G-type, I-type high Ni, I-type low-Ni). We also identified a substantial fraction (22%) of the spherules as being derived from igneous precursors. As these are clearly not related to typical achondrite meteorites as for previously discussed differentiated spherules, we denote these new types of differentiated spherules as D-type spherules. The D-type spherules were classified into four distinct groups. Based on bulk elemental analyses, at least five of the D-type spherules are suggested to be terrestrial in origin, although many others exhibit elemental ratios that are distinct from known planetary bodies and their origins are undetermined. Comparisons of the Australasian microtektites and D-type spherules in the Mg-Fe-Si ternary plot preclude impact spherules as the source of the BeLaU spherules. While there is difficulty in comparing millimeter-sized objects to bulk rock compositions, the coherent elemental abundance patterns we have found suggest magmatic origins rather than nugget effects due to accessory minerals or anthropogenic sources. The chemical compositions of 10 % of the spherules show extremely strong enrichment of refractory lithophile elements such as Be, La and U (BeLaU-type spherules), but very low refractory siderophile elements such as Re. Volatile elements, such as K, Mn, Zn and Pb are very low in the BeLaU-type spherules and were most likely lost by evaporation during passage through the Earth's atmosphere. A magnetite rim is observed in a BeLaU spherule and another D-type spherule, indicating that these went through a rapid cooling process. While the BeLaU-type spherules clearly appear to be derived from material formed by igneous fractionation, their chemical composition is novel in that it is different from known existing solar system materials that have been analyzed thus far, with the KREEP component of the lunar crust being closest. The "BeLaU" samples reflect a highly differentiated, extremely evolved composition, of an unknown origin (see Loeb and MacLeod 2024 for a possible astrophysical origin). Our research team is currently planning another expedition to IM1's site and control regions far from it. Guided by the findings detailed in this paper, the next expedition will retrieve additional materials and address some of the remaining open questions.

These interpretations will be considered critically along with additional results from spherule analysis in future publications. Recovering larger pieces of similar composition from the search area and comparing the findings to distant control regions would certainly help in the interpretation.

*Author contributions*. A. Loeb served as the chief scientist of the expedition, which was coordinated by R. McCallum and funded by C. Hoskinson. All other co-authors were involved in the retrieval and analysis of the spherule samples. The analysis was performed in the laboratories of S. Jacobsen (Harvard University, USA) and R. Tagle (Bruker Corporation, Berlin, Germany).

*Competing interests*. No competing interests.

*Acknowledgements*. We thank C. Hoskinson for funding the expedition, the Galileo Project at Harvard University for administrative and research support and M. Petaev for carrying out EPMA imaging and analyses. We thank the Origins of Life Initiative at Harvard for supporting part of the analytical work on spherules. We are grateful to M. MacLeod, H. Padmanabhan and J. Raymond for helpful comments. We'd also like to thank Matthew J. Genge and another anonymous reviewer for their invaluable and constructive comments throughout the review process which greatly improved the manuscript.

**Supplementary Material.**

Data tables in Excel spreadsheets:
**Table S1**. Major and trace elements concentrations (ppm) measured by micro-XRF for 745 samples (data plotted in **Figures 7, 8, 9, 10a, 11a, 12a**).
**Table S2**. Major elements, Be, Cr, Co, Ni, Sr, La and U concentrations (ppm) measured by ICP-MS for 68 samples (data plotted in **Figures 11b, 12b, 13, 14, 15, 16, 17, 18, 21, 22, 23, 24, 25**).
**Table S3**. Elemental data (ppm) measured by ICP-MS for 12 BeLaU-type spherules and the CI normalizing values of Anders and Grevesse (1989).
**Tables S4–12.** Electron microprobe data (EDS analysis) on the chemical composition in the different regions on the surface of the spherule S21, and polished cross-sections of IS14 SPH4 (S6) and 17NMAG-28 as labeled in **Figures S4–12.**

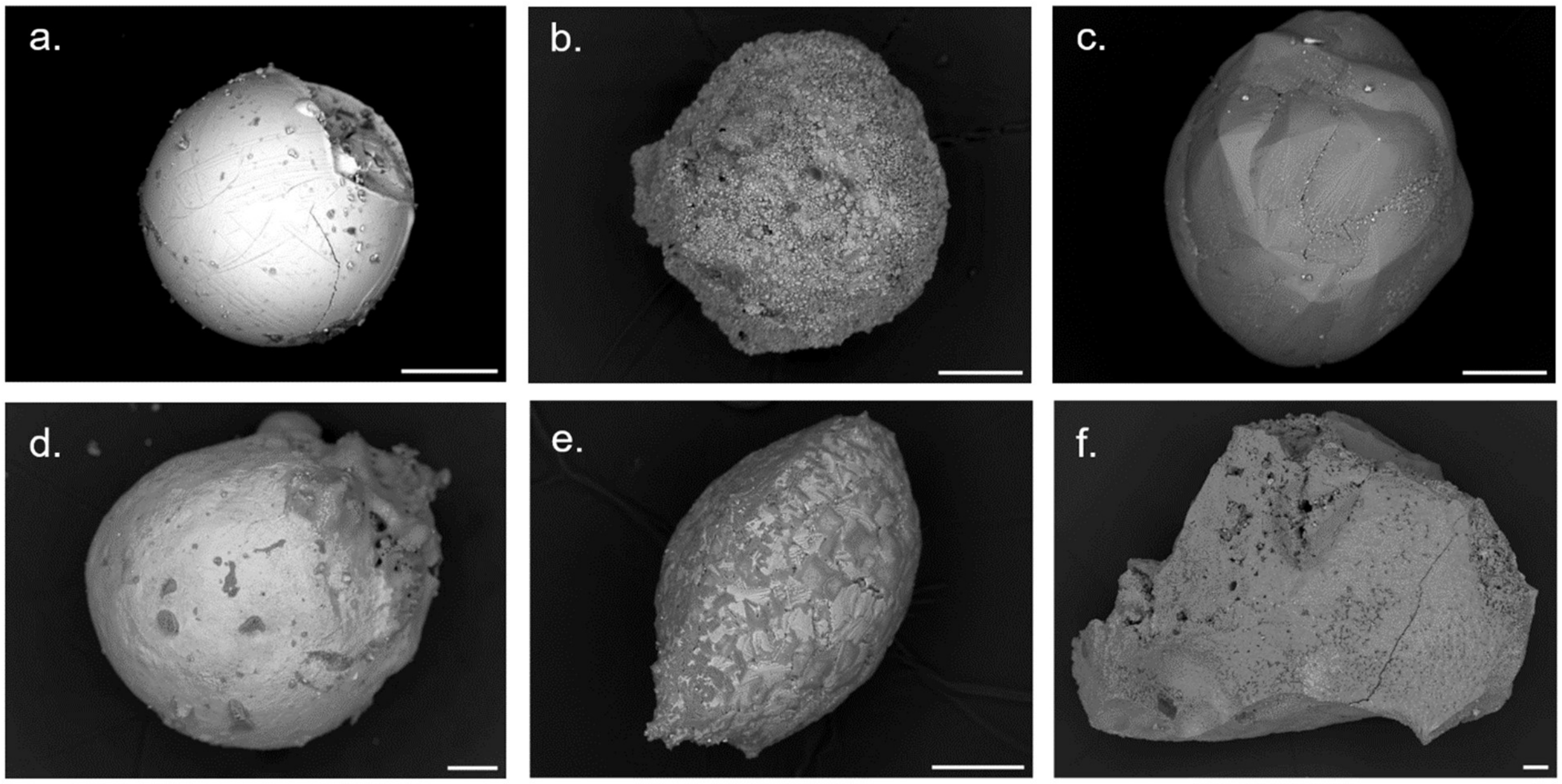

**Figure S1**. BSE images displaying varied morphologies of the samples in this study: a) IS8M2-1 (I-type,

low Ni), b) IS13M1-19 (D-type, low Sr, low Si), c) IS20M-18 (S-type spherule displaying a polyhedral or "turtle-back" morphology), d) IS22M-16 (D-type, high Sr, low Si), e) IS14M(A)-7 (S-type spherule), and f) IS22M1-18 (identified as "shard," no compositional classification). The scale bars indicate 100 microns.

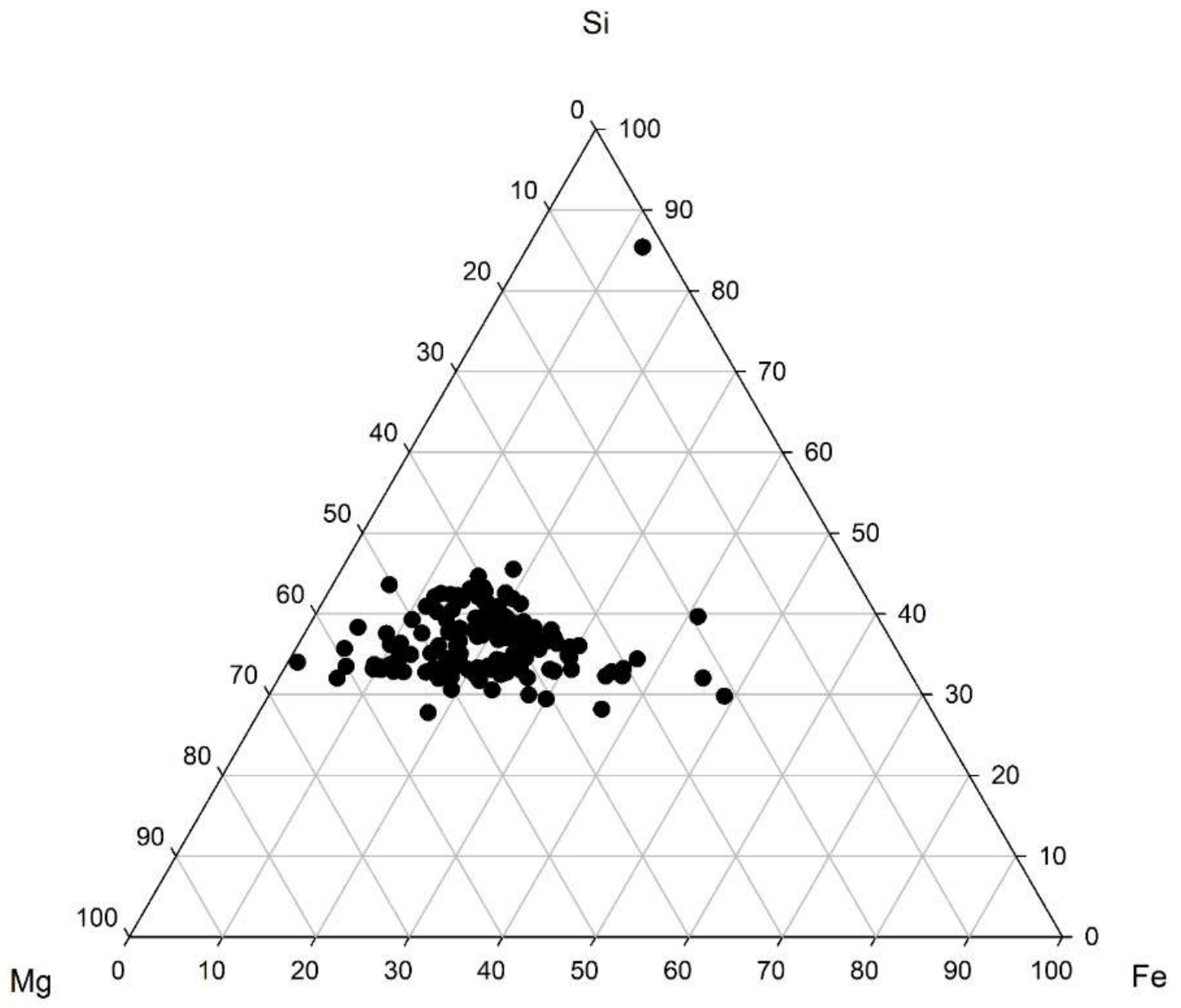

**Figure S2.** Ternary Mg-Si-Fe plot (atom %) of spherules from Rudraswami et al. (2016) normalized to (Mg + Fe + Si) and plotted in elemental proportions. The spherules in this study have been characterized to be S-type spherules. All spherules, except for one, reported by Rudraswami et al. (2016) are S-type spherules, and the one exception does not have trace element enrichments like our BeLaU spherules. Many of the elemental abundances of the BeLaU spherules are 1–2 orders of magnitude more abundant than for the spherules reported by Rudraswami et al. (2016) (their Figure 2).

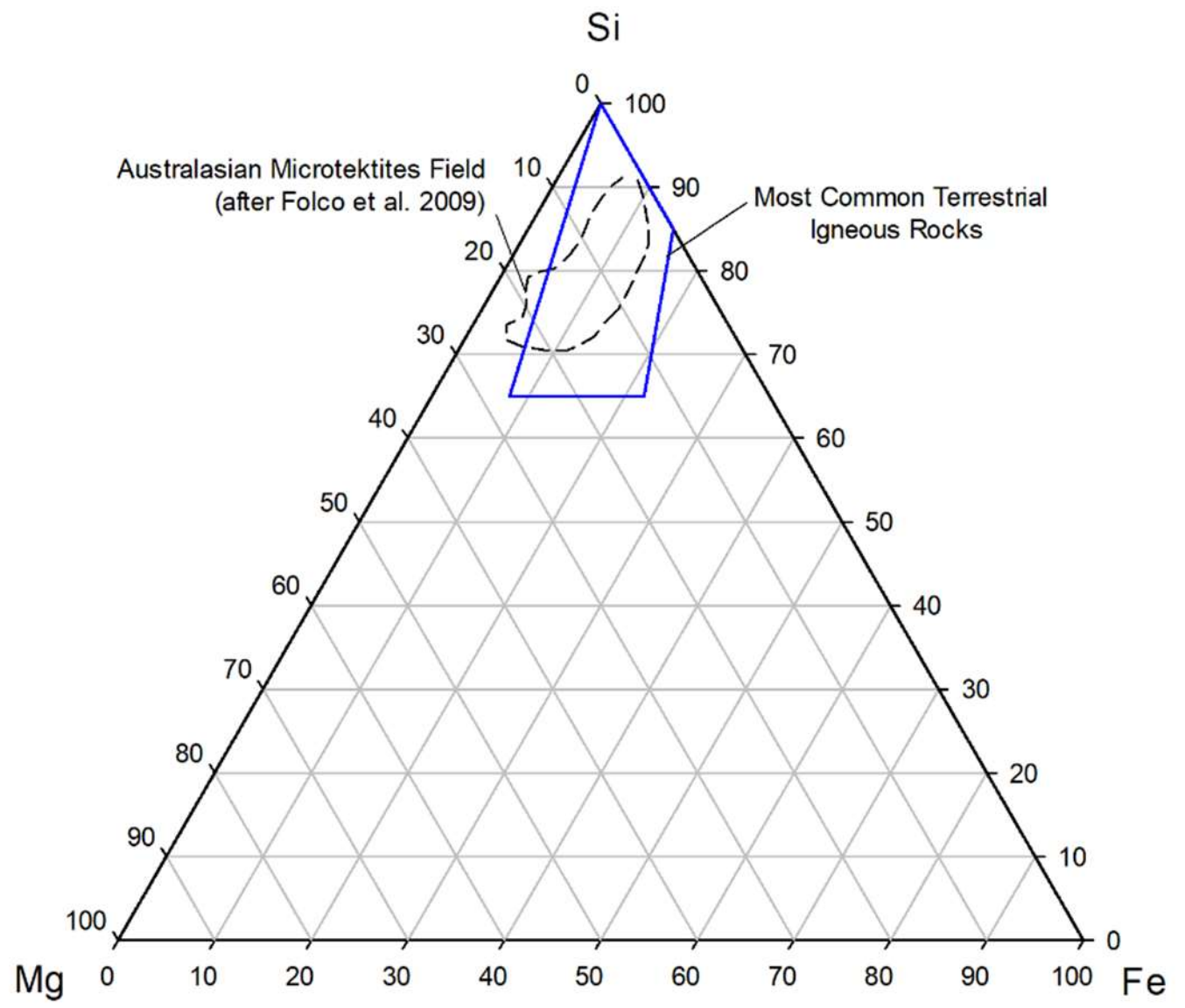

**Figure S3.** Ternary Mg-Si-Fe plot (atom %) delineating the Australasian microtektite field after Folco et al. (2009), which overlaps with the most common terrestrial igneous rocks.

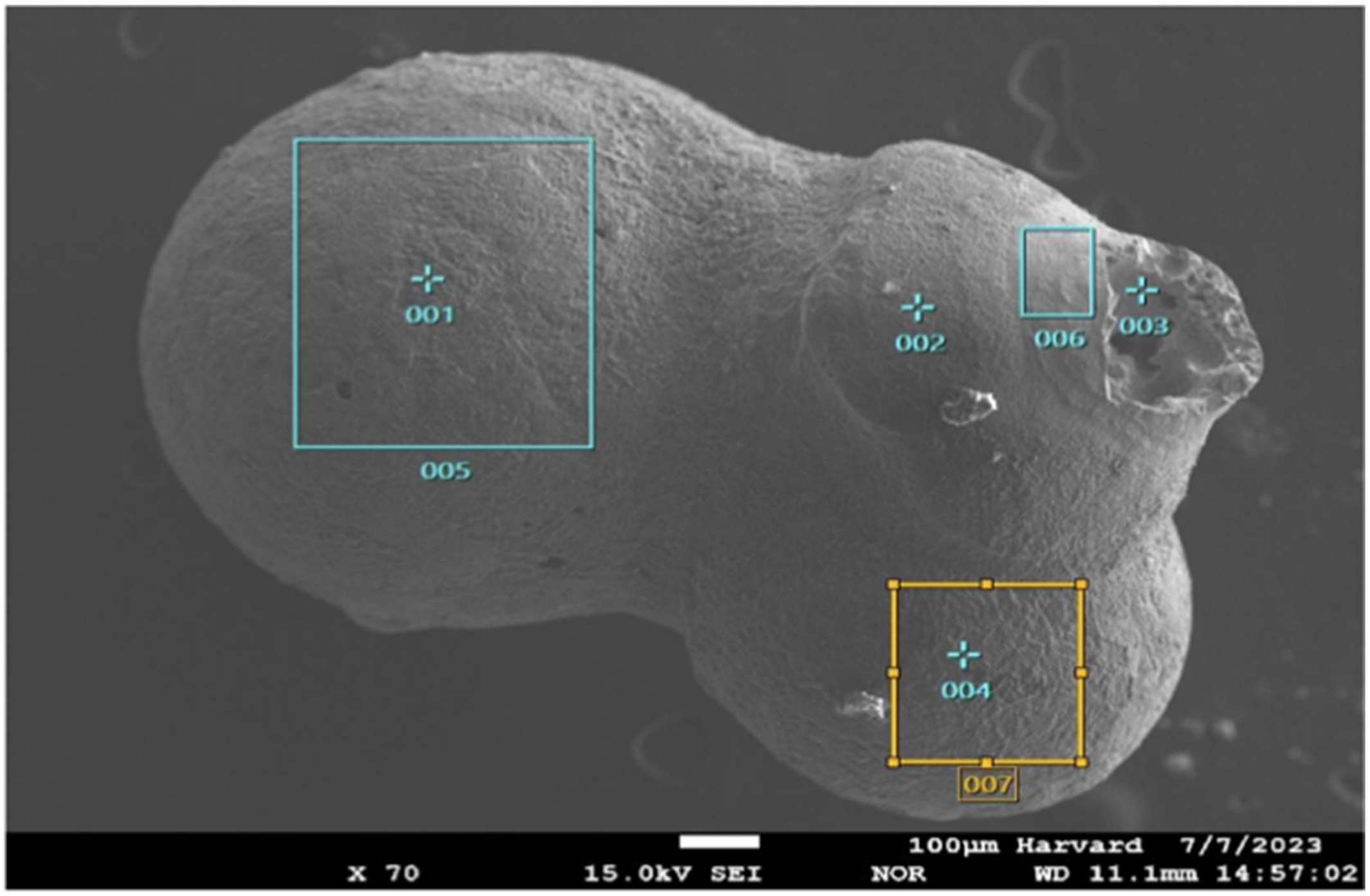

**Figure S4.** Electron microprobe data (EDS analysis) on the chemical composition in the different regions on the surface of the spherule S21. Spot or area analyses are marked with corresponding point numbers in **Table S4**.

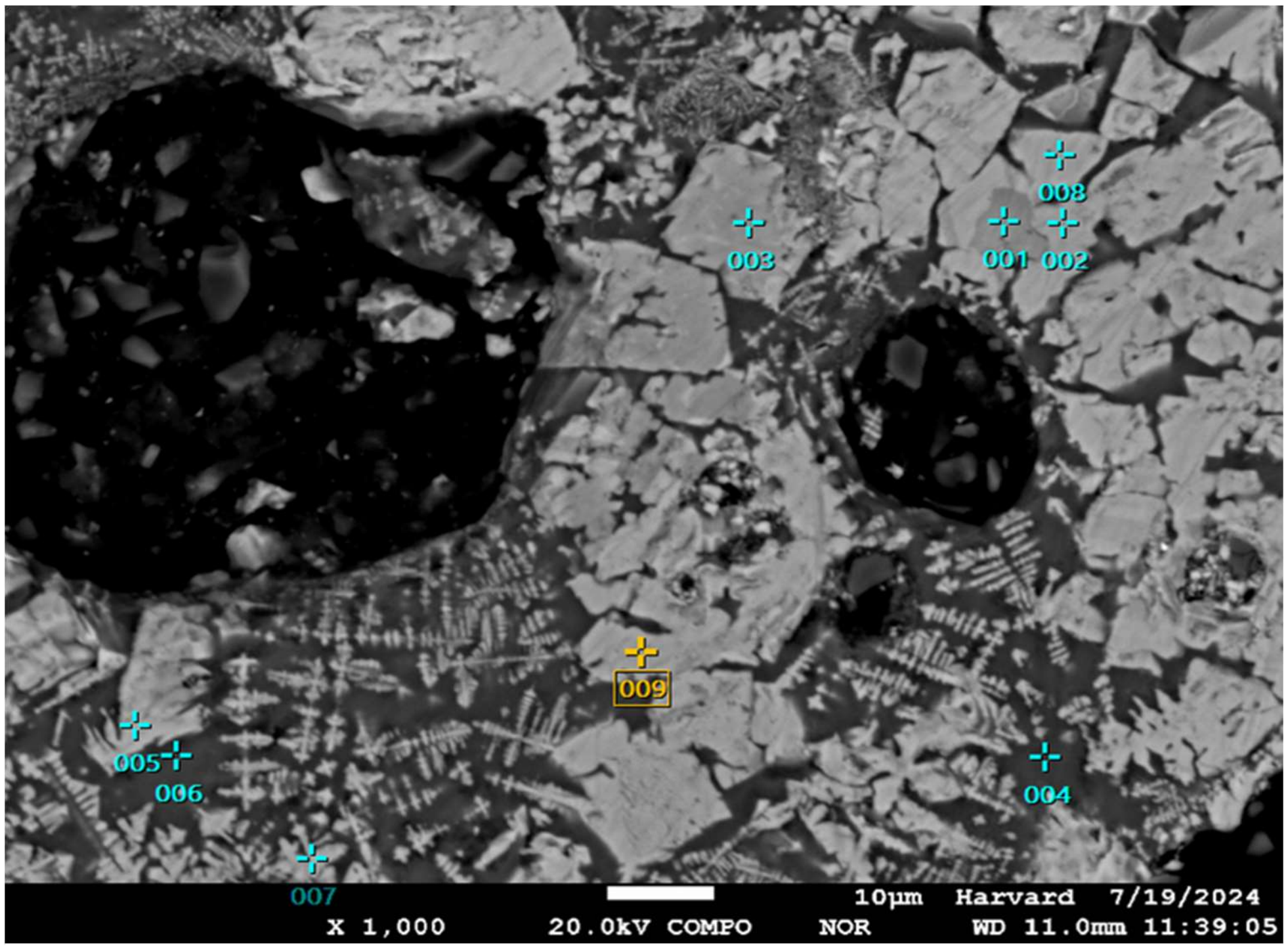

**Figure S5**. Electron microprobe data (EDS analysis) on the chemical composition in the different regions on spherule 17NMAG-28**.** Spot or area analyses are marked with corresponding point numbers in **Table S4**.

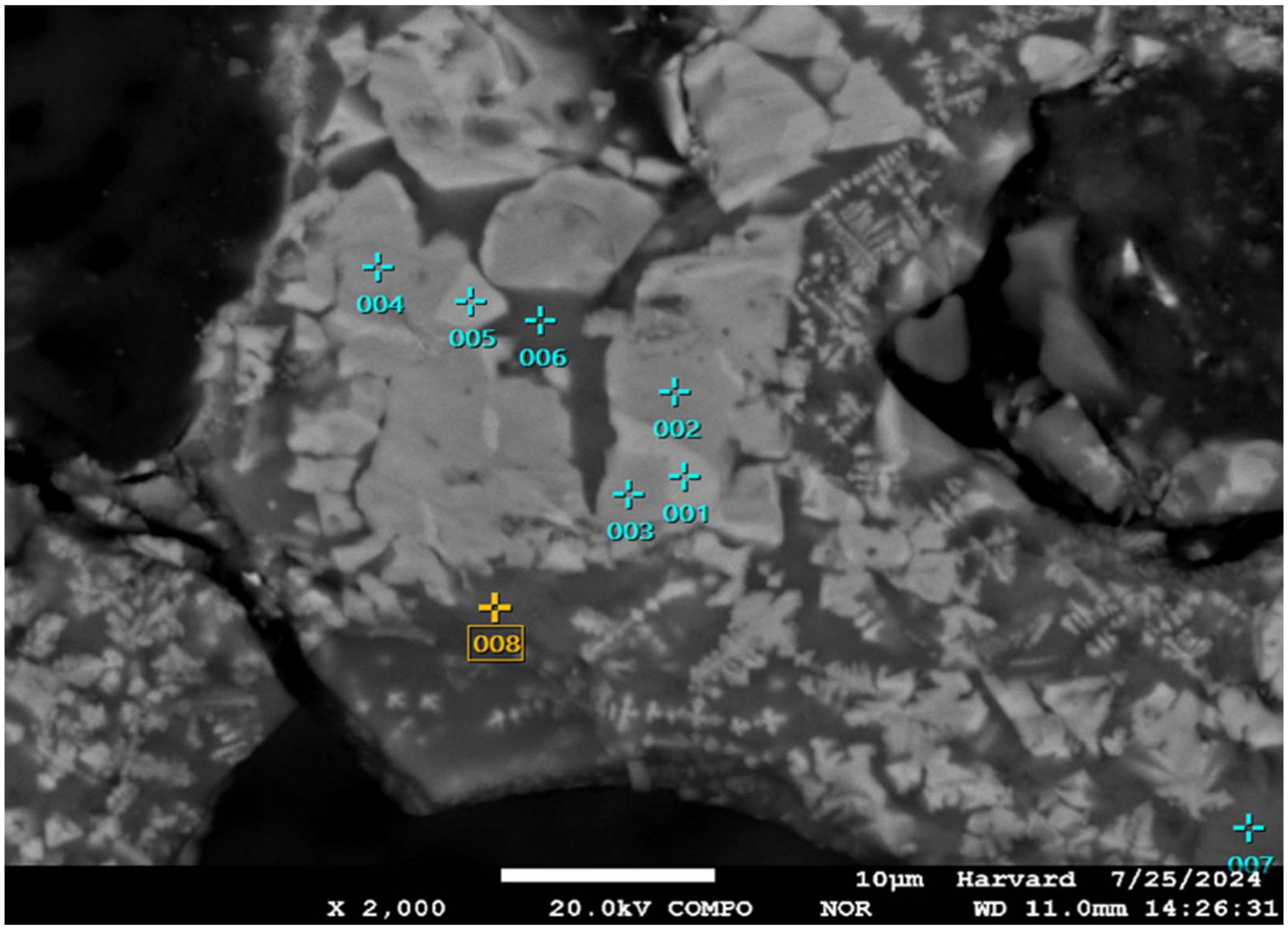

**Figure S6.** Electron microprobe data (EDS analysis) on the chemical composition in the different regions on spherule 17NMAG-28**.** Spot or area analyses are marked with corresponding point numbers in **Table S4**.

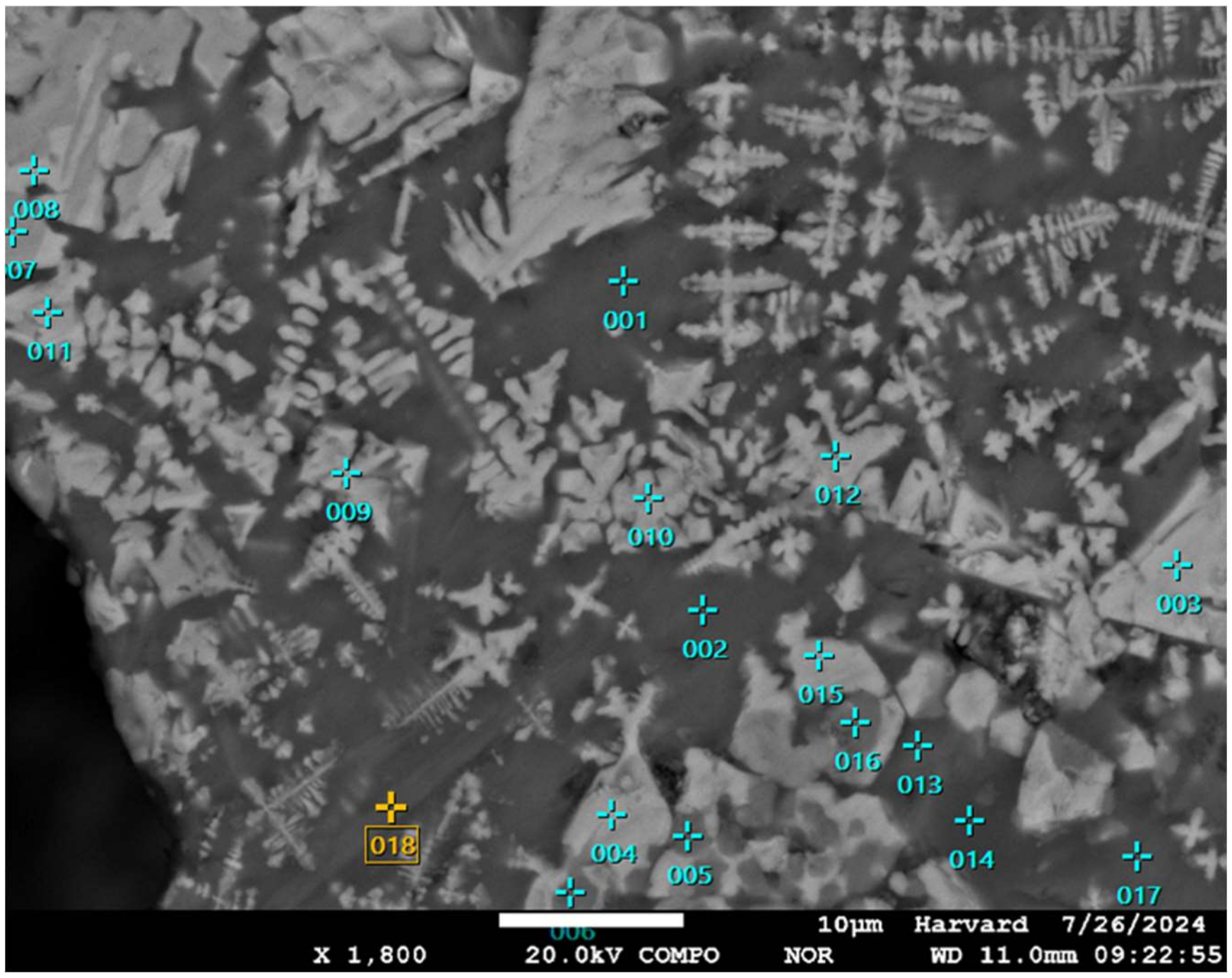

**Figure S7.** Electron microprobe data (EDS analysis) on the chemical composition in the different regions on spherule 17NMAG-28**.** Spot or area analyses are marked with corresponding point numbers in **Table S4**.

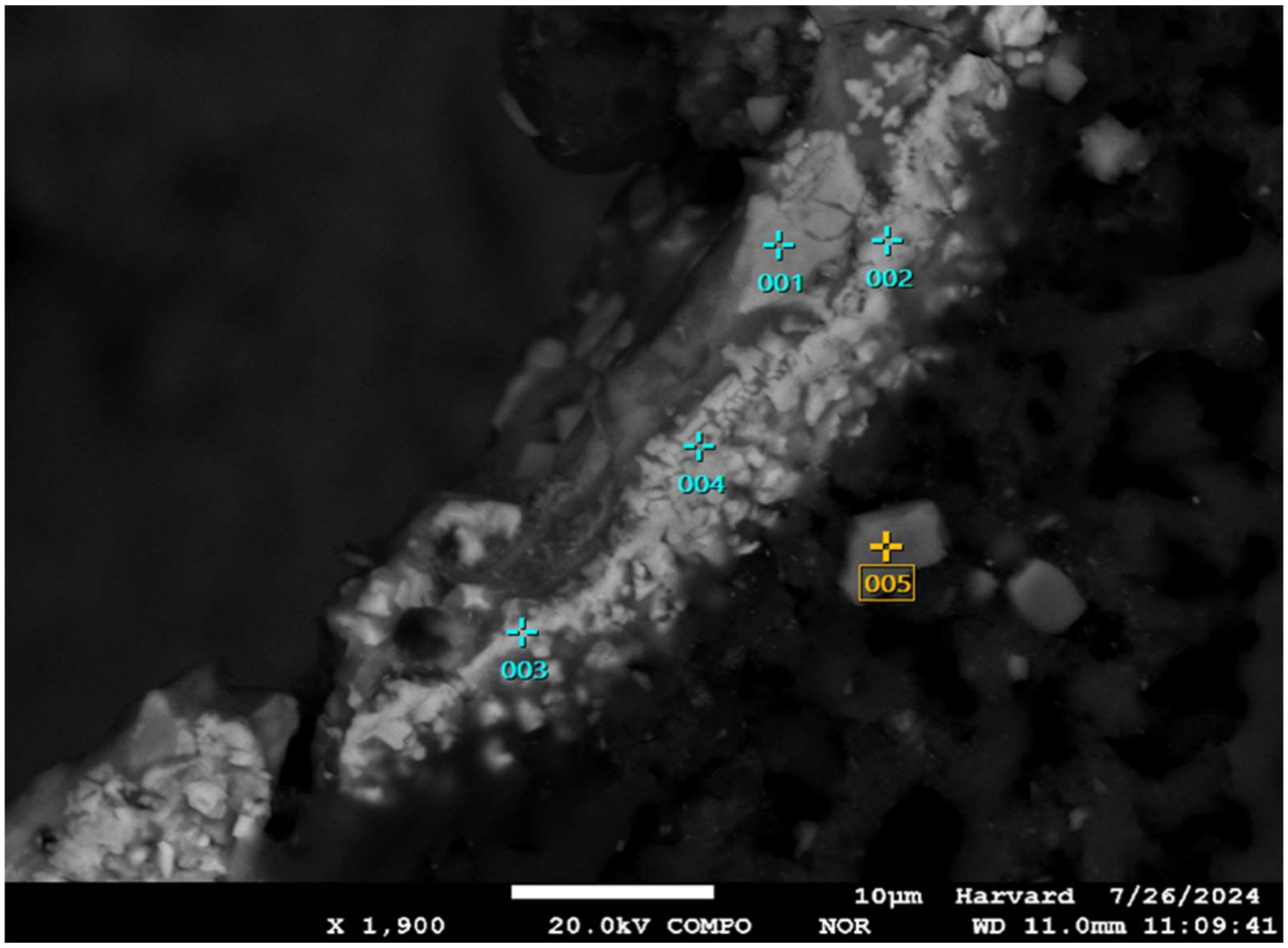

**Figure S8.** Electron microprobe data (EDS analysis) on the chemical composition in the different regions on spherule 17NMAG-28**.** Spot or area analyses are marked with corresponding point numbers in **Table S4**.

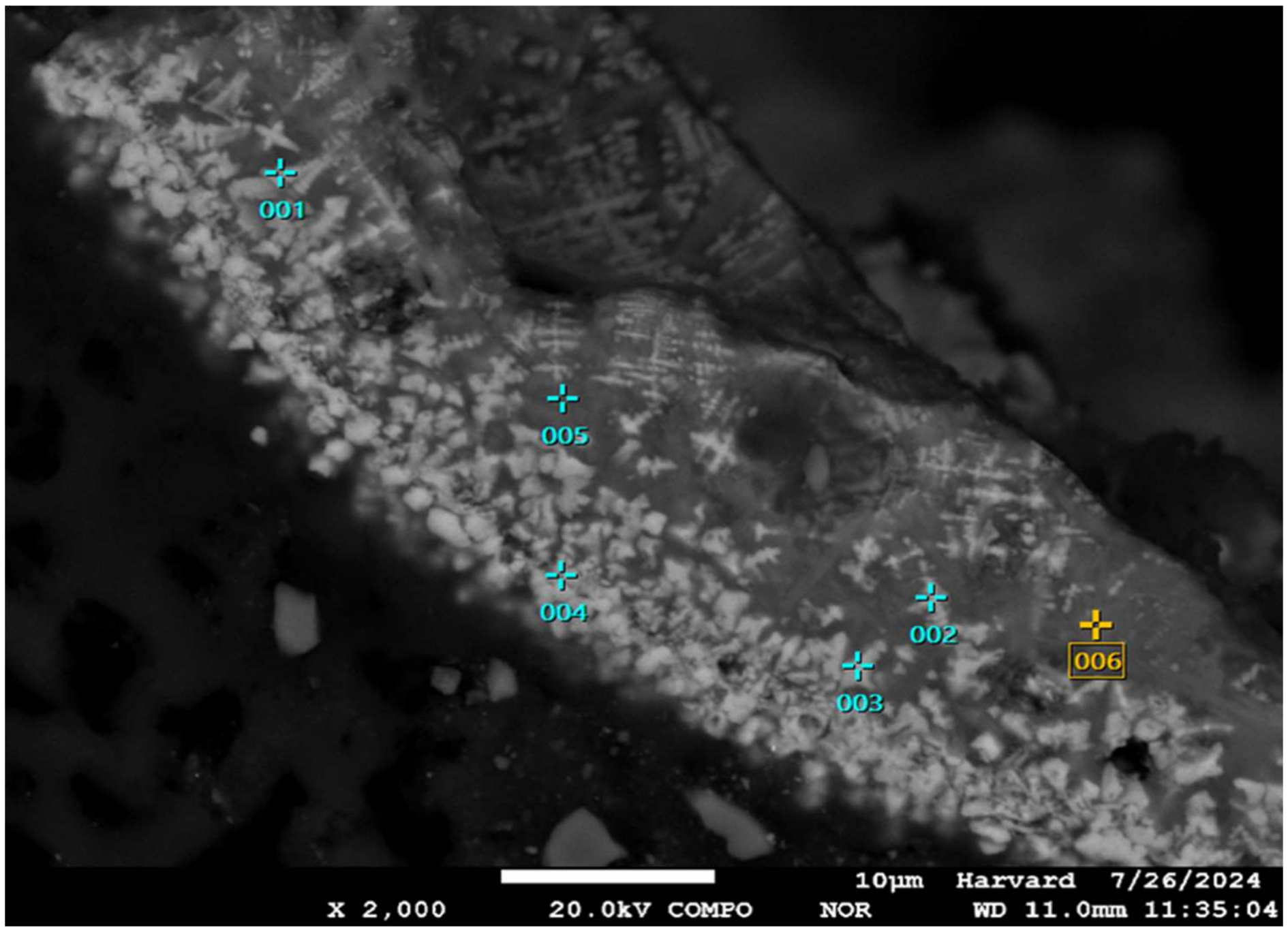

**Figure S9.** Electron microprobe data (EDS analysis) on the chemical composition in the different regions on spherule 17NMAG-28**.** Spot or area analyses are marked with corresponding point numbers in **Table S4**.

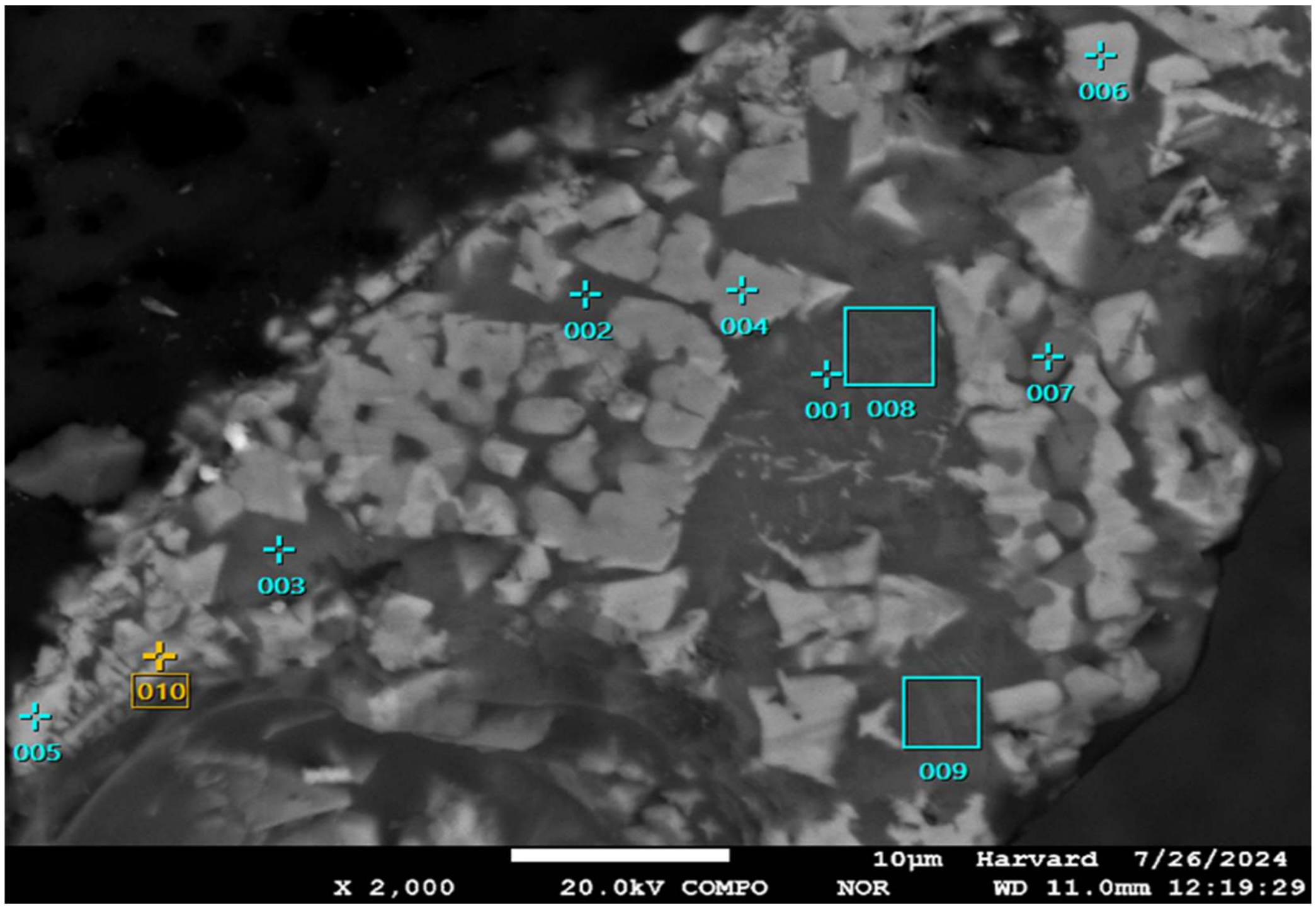

**Figure S10.** Electron microprobe data (EDS analysis) on the chemical composition in the different regions on spherule 17NMAG-28**.** Spot or area analyses are marked with corresponding point numbers in **Table S4**.

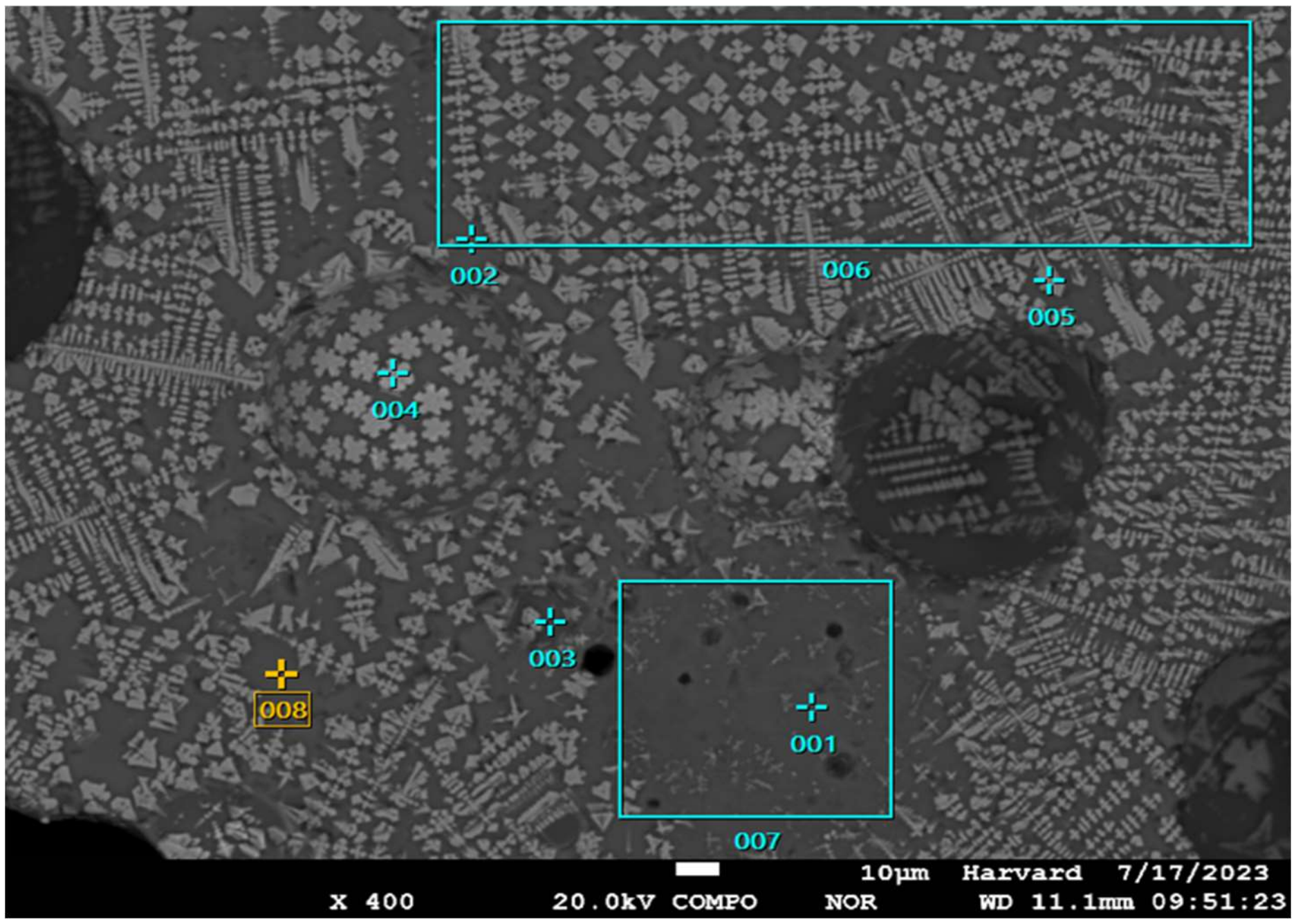

**Figure S11.** Electron microprobe data (EDS analysis) on the chemical composition in the different regions on spherule IS14 SPH4**.** Spot or area analyses are marked with corresponding point numbers in **Table S4**.

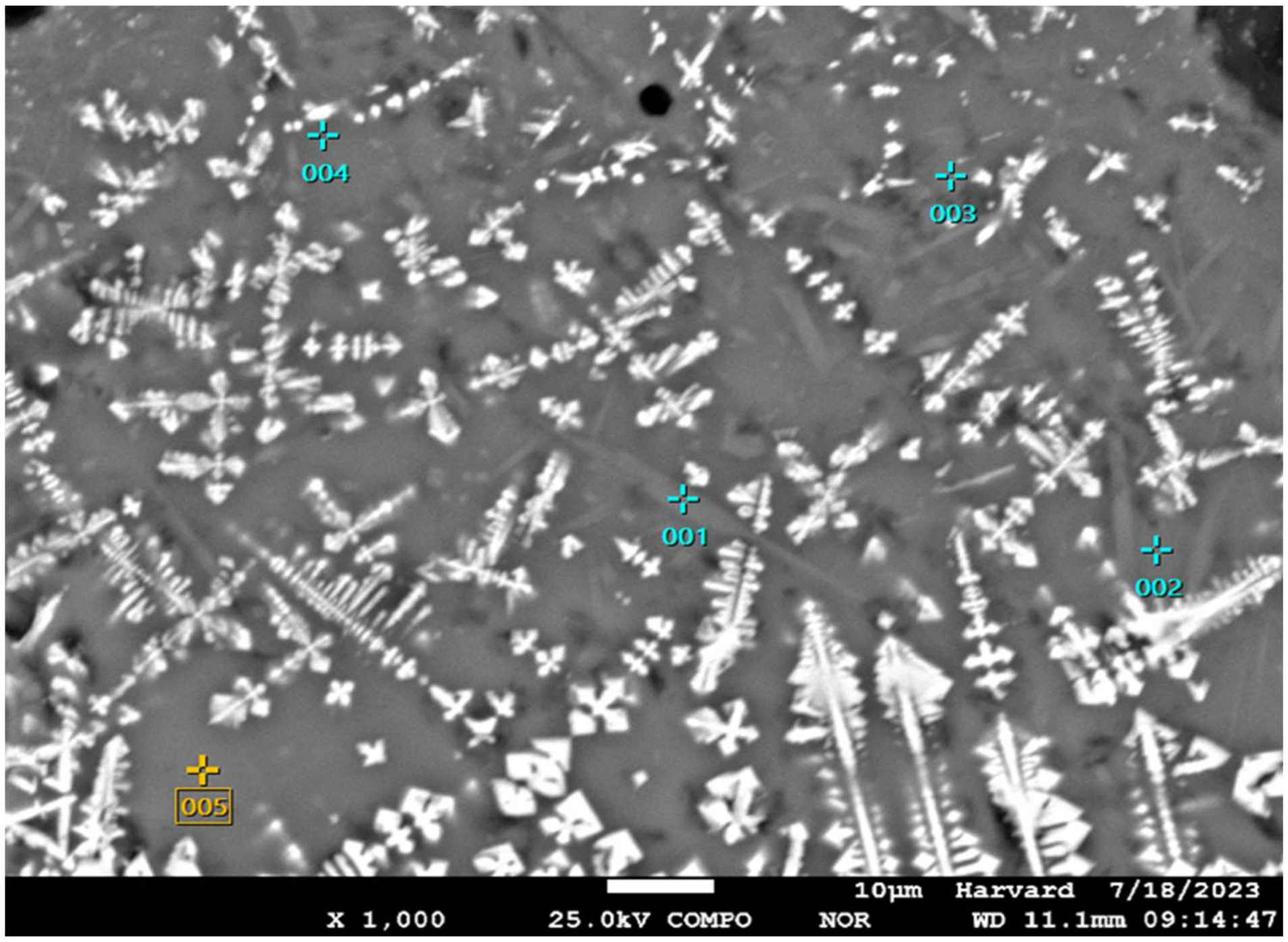

**Figure S12**. Electron microprobe data (EDS analysis) on the chemical composition in the different regions on spherule IS14 SPH4 (S6)**.** Spot or area analyses are marked with corresponding point numbers in **Table S4**.